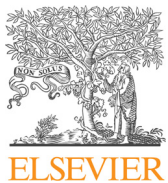
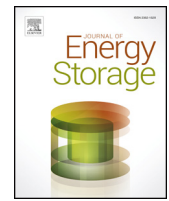

# Data-driven nonparametric Li-ion battery ageing model aiming at learning from real operation data - Part B: Cycling operation

M. Lucu[a,b,*], E. Martinez-Laserna[a], I. Gandiaga[a], K. Liu[c], H. Camblong[b,d], W.D. Widanage[c], J. Marco[c]

[a] *Ikerlan Technology Research Centre, Basque Research and Technology Alliance (BRTA). Pº J.M. Arizmendiarrieta, 2. 20500 Arrasate-Mondragón, Spain*
[b] *University of the Basque Country (UPV/EHU), Department of Systems Engineering & Control. Europa Plaza, 1. 20018 Donostia-San Sebastian, Spain*
[c] *WMG, University of Warwick, Coventry, CV4 7AL, UK*
[d] *ESTIA Research, Ecole Supérieur des Technologies Industrielles Avancées (ESTIA), Technopole Izarbel, 64210 Bidart, France*

ABSTRACT

Conventional Li-ion battery ageing models, such as electrochemical, semi-empirical and empirical models, require a significant amount of time and experimental resources to provide accurate predictions under realistic operating conditions. At the same time, there is significant interest from industry in the introduction of new data collection telemetry technology. This implies the forthcoming availability of a significant amount of real-world battery operation data. In this context, the development of ageing models able to learn from in-field battery operation data is an interesting solution to mitigate the need for exhaustive laboratory testing.

In a series of two papers, a data-driven ageing model is developed for Li-ion batteries under the Gaussian Process framework. A special emphasis is placed on illustrating the ability of the Gaussian Process model to learn from new data observations, providing more accurate and confident predictions, and extending the operating window of the model.

The first paper of the series focussed on the systematic modelling and experimental verification of cell degradation through calendar ageing. Conversantly, this second paper addresses the same research challenge when the cell is electrically cycled. A specific covariance function is composed, tailored for use in a battery ageing application. Over an extensive dataset involving 124 cells tested during more than three years, different training possibilities are contemplated in order to quantify the minimal number of laboratory tests required for the design of an accurate ageing model. A model trained with only 26 tested cells achieves an overall mean-absolute-error of 1.04% in the capacity curve prediction, after being validated under a broad window of both dynamic and static cycling temperatures, Depth-of-Discharge, middle-SOC, charging and discharging C-rates.

## 1. Introduction

Lithium-ion (Li-ion) battery technology has gained a significant market share as the principal energy storage solution for many industrial applications, mainly due to its high energy efficiency and high specific energy and power [1,2]. However, Li-ion batteries are still relatively expensive compared to other storage technologies, and their performance is known to decline over time and use, which threatens their competitiveness against more affordable solutions [2,3]. As highlighted in the first paper of the series [4], the development of accurate battery ageing models could play a key role to overcome such barriers; however, the obtention of accurate models typically requires a high number of laboratory tests.

As suggested in a previous publication, a suitable solution to reduce the number of laboratory tests could be the development of ageing models capable to continuously learn from streaming data [5]. Following this approach, reduced laboratory tests could be used to develop a preliminary ageing model. Further, once the battery pack has been implemented and deployed, in-field data extracted by the data acquisition system could allow updating the preliminary ageing model. In this way, the ageing model would be continuously upgraded, improving prediction accuracy, extending the operating window of the model itself and providing useful information for predictive maintenance, adaptive energy management strategies or business case redefinition.

* Corresponding author: Ikerlan Technology Research Centre, Basque Research and Technology Alliance (BRTA), Pº J.M. Arizmendiarrieta, 2. 20500 Arrasate-Mondragón, Spain.
*E-mail address:* mlucu@ikerlan.es (M. Lucu).

In a previous study, a critical review on self-adaptive ageing models for Li-ion batteries was presented, in which the Gaussian Process (GP) method was identified as the most promising candidate [5]. In fact, beyond their ability to perform probabilistic, relatively robust and computationally acceptable predictions, these models enjoy the very interesting advantage of being nonparametric: in other words, the complexity of these models depends on the volume of training data. Within the context of Li-ion ageing prediction, this implies:

- *A progressive spread of the operating window for the model.* Each time a new data sample related to previously unobserved operating conditions is included within the training set, additional knowledge is obtained about the influence of stress-factors on ageing. The resulting models should provide an increasingly comprehensive picture of the ageing of Li-ion batteries.
- *A higher level of specialisation of the model.* The preliminary ageing model developed from the laboratory ageing data could be upgraded by including new training data extracted from the in-field operation. In-field data encodes the intrinsic operating profiles of each application, as well as the corresponding battery ageing. This implies the possibility to move from a generic ageing model to a specialised model tailored to the specific applications.

Each time input values are presented to the model to perform a prediction, the GP model retrieves similar data samples in the training dataset to produce analogous predictions. A continuously fed training dataset implies an increased number of similar data, allowing more accurate and confident predictions.

From a broader perspective, most of the data-driven ageing models proposed in the literature refers to the degradation of the Li-ion batteries when the cell is electrically cycled. However, almost all of them are developed based on cycling profiles corresponding to specific applications, and do not consider the ageing of Li-ion batteries from a general prospect. Accordingly, the most critical gaps identified in the literature regarding data-driven Li-ion ageing models are i) the under-utilisation of key predictive features (e.g. values of the different stress-factors) and ii) the insufficient validation of the proposed models [5]. These gaps strongly limit the accuracy and applicability of the models within the context of real deployment. In this sense, investigation in data-driven Li-ion ageing models should be more focussed on the implementation or discovery of features presenting strong predictive capabilities (as suggested in [6]), as well as the deeper validation of the developed models under broad operating conditions.

The GP framework has already been introduced for Li-ion battery ageing predictions [4,7–15]. The present study aims to extend existing research by integrating the following main contributions:

i) The development of a generic data-driven cycle ageing model, able to perform accurate capacity loss predictions for a broad range of cycling conditions, and usable for a large diversity of Li-ion battery applications.
ii) The extension and validation of the main contributions introduced in the first paper of the series in the context of calendar ageing, bringing them to the battery cycling use-case. Thus, the ability of GP models to learn from new data is analysed, illustrating their capability to provide more accurate and confident ageing predictions when integrating previously unobserved operating conditions, extending this way the operating window of the model. Furthermore, a compositional covariance function is introduced, tailored to Li-ion battery cycle ageing prediction.

Additionally, this study also extends the secondary contributions presented in the first paper of the series to the cycle ageing use-case:

i) The quantification of the minimal number of laboratory tests required for the design of an accurate cycle ageing model for a broad operating window.
ii) The validation of the proposed ageing model with an extensive experimental ageing dataset, involving 122 cells tested during more than three years at static conditions, and 2 additional cells tested at dynamic operating conditions.
iii) The sensitivity analysis of the capacity loss with respect to the different stress-factors, from the point of view of the developed model. As explained in this paper, the developed covariance function shares the particularity of quantifying the relevance of each input variable for predicting the defined output variable. This could provide some intuitions about e.g. which stress-factors are the most impactful on the capacity loss, producing useful insights for the design of energy management strategies. Such analysis was not performed in the field of Li-ion battery ageing prediction.

The body of the research undertaken is presented through a two-part series. The first paper focussed on the systematic modelling and experimental verification of cell degradation through calendar ageing [4]. Conversantly, this paper addresses the same research challenge when the cell is electrically cycled. During many real-world conditions, the cell will be subject to both calendar and cyclic ageing. The relative importance of each will be highly dependent on the nature of the use-case. The integration of both forms of ageing, within the context of defining a holistic view of lithium-ion degradation modelling is a challenging research task, discussed further within [16,17] and is the subject of ongoing research by the authors further extending the research presented here and in [4].

This paper is structured as follows, Section 2 describes the experimental ageing tests carried out in order to produce the ageing data. The raw data obtained from the experimental tests are analysed and processed before the development of the model. Section 3 details the processing of the raw data and evaluate the relevance of the obtained data for ageing modelling. Section 4 introduces the general background of the GP theory, and Section 5 presents the development of the proposed cycle ageing model under the GP framework. In Sections 6 and 7, the prediction results of the developed model are presented for the cells cycled at static and dynamic operating conditions, respectively. Furthermore, both sections aim to illustrate the ability of the GP model to learn from new data observation. Section 8 discusses the obtained results, leading to the identification of the limitations of the study and opening the way to further works. Finally, Section 9 closes the study depicting the main conclusions.

## 2. Experimental cycle ageing data

Within the context of the European project titled as Batteries2020, extensive experimental works were carried out over a time span of more than three years, in order to analyse the ageing of Li-ion batteries, covering different possible operations. The capacity retention of a 20 Ah Lithium Nickel-Manganese-Cobalt (NMC 4:4:2) cathode-based pouch cell with a graphite anode was evaluated. The nominal characteristics of the cell, the operating conditions recommended by the manufacturer, as well as the experimental results obtained from a testing batch of 32 cells related to the study of the ageing in storage operation, were described in the first paper of the series [4]. In this second paper, the experimental works associated with the study of the cycling operation will be presented.

From the ageing point of view, the operation of a Li-ion battery in cycling is conditioned by the level of different stress-factors, mainly identified in the literature as the operating temperature, Depth-Of-Discharge (DOD), average State-Of-Charge (middle-SOC), and the charging and discharging C-rates [18]. A total of 124 cells were cycled in temperature-controlled climatic chambers, at different combinations of such stress-factors. Periodical characterisation tests were carried out at 25°C in order to evaluate the progressive capacity retention of the cells. The determination of the capacity started 30 min after its surface temperature reached 25°C degrees, ensuring that the cells has stabilised at the target temperature. The test started with a constant current – constant voltage (CC-CV) charge: the CC charge was done at 6.667 A (C/3) until reaching 4.15 V, and the following CV charge was stopped when achieving current values below 1 A (C/20). After a period of 30 min, the cell was discharged using a CC discharge current at 6.667 A (C/3) until the terminal voltage measured 3 V, followed by a pause period of 30 min. The procedure was repeated three times. The capacity value obtained in the last repetition was considered as the cell capacity.

Depending on the variability of the stress-factors' profiles in the whole duration of the tests, two types of ageing experiments were distinguished, namely i) the ageing tests at static operating conditions and ii) the ageing tests at dynamic operating conditions.

### 2.1. Experimental ageing tests at static operating conditions

In the ageing tests performed at static conditions, the value of the stress-factors remained constant throughout the whole duration of the tests. 122 cells were tested at 34 different operating conditions, specified in Table A1, Appendix A. Most of these tests were performed in the laboratories of the Vrije Universiteit Brussel and were completed by the laboratories of Ikerlan Technology Research Centre, ISEA – RWTH, Leclanché and Centro Ricerche Fiat. The cells were characterised every 4000 Ah, or equivalently 100 Full-Equivalent-Cycles (FECs). In order to ensure the repeatability of the results, at least 3 cells were allocated to each testing condition. The capacity curves resulting from the experimental ageing tests at static conditions are observable in Fig. B.1, Appendix B. The variability of the capacity curves obtained for each tested cycling conditions is indicated in Table B1, Appendix B.

### 2.2. Experimental ageing tests at dynamic operating conditions

As the battery stress conditions in real-world applications are not constant over time, the developed ageing models should be able to perform accurate predictions at dynamic operating profiles. The ability of the GP model to learn from dynamic profiles should also be analysed. Therefore, 2 additional cells were tested at dynamic profiles of the different stress-factors. The value of the stress-factors was modified between each characterisation test, during the whole duration of the tests. One cell was tested at constant 80% DOD, 50% middle-SOC, C/3 rate in charge, 1C rate in discharge and a variable temperature profile following the seasonal temperatures over a year, between a range of 15°C–36°C. Furthermore, one additional cell was submitted to the same seasonal temperature profile, but also variable DOD, middle-SOC and charging and discharging C-rates. The cells were characterised approximately every 28 days. The variation profiles of the stress-factors, as well as the corresponding capacity retention of the tested cells are depicted in Fig. B.2, Appendix B.

### 3. Data preprocessing and evaluation of the resulting data

In the context of data-driven modelling, it is important to analyse and preprocess the raw data before any modelling task, in order to address data inconsistency and noise issues and achieve effective models [19]. The capacity curves obtained from the experimental ageing tests described in Section 2 present clearly three distinct phases, as illustrated in Fig. 1.

The first phase corresponds to an initial capacity rise appearing at the Beginning Of Life (BOL). As detailed in the first paper of the series, this behaviour could be explained by the geometrical characteristics of the cells and it is not related to any ageing mechanism. Accordingly, the data corresponding to Phase 1 was discarded for the development of the ageing model. During the data preprocessing stage, the maximal capacity point of each cell was designated as the BOL point and assigned to the 'zero cycled Ah-throughputs' state. The second phase is characterised by a progressive rate-constant decrease of the cell capacity, and it is sometimes followed by a third phase describing a sudden capacity drop, as illustratively depicted in Fig. 1. According to [20], in these tested cells, this third phase was provoked by the occurrence of lithium plating. For the reasons exposed in the first paper of the series, the modelling of the Phase 3 remained out of the scope of the study, and the corresponding data was discarded from the modelling dataset.

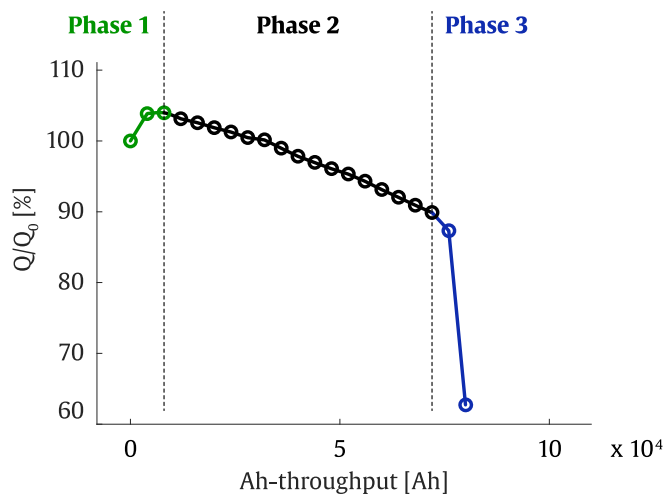

**Fig. 1.** The three different phases of the capacity retention curve of the cells. The first phase is an increase of the capacity, the second is progressive degradation and third phase is a sudden capacity drop. Modified from [20].

Therefore, in the context of this study, the modelling work focussed on capturing the relations between the cycling conditions and the capacity loss of the cells, during the progressive degradation corresponding to the second phase in Fig. 1.

Besides, some unexpected trends were identified within the experimental data, for instance, abnormally reduced capacity measurements around 25000 Ah in the cells #21-23 (green curves in Fig. B.1(a), Appendix B). Such deviations are related to procedural errors during the capacity tests (e.g. exchange of the testing device, etc.). These noisy data samples could affect the performances of the model and were therefore removed from the modelling dataset. Furthermore, the cell #56 showed a clearly defective behaviour (isolated red curve in Fig. B.1(f), Appendix B) and was also discarded from the dataset.

On average, 76.5% of the raw experimental data corresponding to the static ageing conditions was preserved after the preprocessing stage. The percentage of the remaining data for each cell is indicated in Table 1. Overall, all the ageing conditions of the initial experimental ageing matrix were still represented in the processed dataset. It is noteworthy that most of the discarded data corresponds to cells cycled at low DOD values, due to the decision to neglect the initial capacity rise points. Regarding to the cells submitted to dynamical ageing profiles, 90% and 95.45% of the ageing data was maintained for the cells #124 and #125 respectively. Figs. 2 and 3 illustrates the resultant ageing data obtained after the preprocessing stage.

The analysis of the capacity curves in Fig. 2 allows to understand the relations between the values of the different stress-factors and the underlying ageing of the cells. Comparing the curves corresponding to identical DOD operations in Fig. 2(a), (b) and (c), it is noteworthy that the increased cycling temperature in (c) accelerated the capacity loss of the cells. This observation is in accordance with the literature [21,22]. In fact, the growth of the SEI layer is a chemical reaction and then obeys to the Arrhenius law: the SEI formation rate increases exponentially with temperature.

By studying them independently, the Fig. 2(a), (b) and (c) also illustrate the DOD dependency of the capacity loss. Higher values of the DOD increased the capacity loss. As explained in [23], at relatively low current rates of battery operation the SEI cracking and reforming is the main mechanism inducing capacity loss. Such capacity loss was shown to be dependent to the state of lithiation swing (which could be approximated by the DOD) of the electrode, during lithiation [23]. Similar experimental results were reported in [24–26].

**Table 1**
Remaining data percentage ranges for each cycling condition, after the data preprocessing.

| DoD [%] | MidSOC [%] | 25 °C C/3-1C | 25 °C 1C-1C | 35 °C C/3-C/3 | 35 °C C/3-2C | 35 °C C/2-1C | 35 °C C/3-1C | 35 °C 1C-1C | 35 °C 2C-1C | 35 °C 2C-2C | 45 °C C/3-1C |
|---|---|---|---|---|---|---|---|---|---|---|---|
| 100 | 50 | 89.6% | | | | | 85.7 – 92.8% | | | | 83.3 – 91.6% |
| 80 | 50 | 70 – 95% | 88.5 – 94.1% | | | | 62.5 – 100% | | | | 80 – 92.8% |
| 65 | 50 | 50 – 87.5% | | | | | 88.8 – 100% | | | | 80 – 100% |
| 50 | 65 | | | | | | 88.2 – 100% | | | | |
| 50 | 50 | 82.7 – 72.4% | | 66.6 – 83.3% | 77.2 – 81.8% | 100% | 88.2 – 100% | 66.6 – 88.8% | 50 – 66.6% | 62.6 – 72.7% | 73.3 – 80% |
| 50 | 35 | | | | | | 76.4 – 77.7% | | | | |
| 50 | 80 | | | | | | 77.7 – 88.8% | | | | |
| 35 | 50 | 50% | | | | | 83.3 – 94.4% | | | | 80% |
| 20 | 65 | | | | | | 33.3 – 44.4% | | | | |
| 20 | 80 | | | | | | 55.5 – 61.1% | | | | |
| 20 | 35 | 47 – 73.3% | | | | | 11.1 – 44.4% | | | | 63.6 – 78.5% |
| 20 | 20 | | | | | | 38.8 – 50% | | | | |
| 10 | 80 | | | | | | 94.4% | | | | |
| 10 | 65 | | | | | | 33.3% | | | | |
| 10 | 20 | | | | | | 22.2 – 66.6% | | | | |

Regarding the effect of the middle-SOC stress-factor, the cycling at higher lithiation ranges of the anode is expected to lead to accelerated ageing, due to i) the effect of the calendar ageing, in which higher SOC values induce faster degradation [20,27], ii) the increased mechanical stress accumulated in the anode at higher lithiation states, conducting to accentuated SEI cracking and reforming [28] and iii) the crossing of the transitions between voltage plateaus of the negative electrode, which provokes changes in the lattice parameters of the material and leads to material expansion and contraction, increasing again the mechanical stress [26]. The latter element suggests a U-shape dependency of the capacity loss to the middle-SOC, with an optimum around 50% SOC and stronger degradations at higher and lower cycle ranges [26]. A similar behaviour is observable in Fig. 2(f), in which the cells cycled at 35% and 65% middle-SOC aged slightly faster than the cells operating at 50% middle-SOC. As for the 10% and 20% DOD operation, Fig. 2(d) and (e) reflect an increased capacity loss at 80% middle-SOC operation, compared with lower middle-SOC levels.

Furthermore, the effect of high charging and discharging C-rate values on capacity loss was also demonstrated in the literature. High C-rates lead to additional stress in the electrodes, due to i) a non-homogeneous intercalation of lithium on graphite which create Li-concentration gradients and ii) more important volume expansions and compressions [28]. This increase the probability of particle fracture, conducting to a loss of active material. In the negative electrode, the particle cracking reveals fresh anode surface, which react with electrolyte reforming SEI and augmenting capacity loss [28,29]. Furthermore, the charging at high C-rate could generate the lithium plating reaction, because of the heterogeneous lithium repartition in the material which could locally induce voltages close to the 0 V vs. Li/Li$^+$ [30]. The study of the C-rate effect in the experimental works was limited to a C/3–2C range, in order to obtain enough resolution. These are relatively low levels compared to the actual EV market requirement (~6C in charge [31]). The obtained results showed relatively high variability, and not clear influence of the C-rate was remarkable below 1C for both charging and discharging C-rates (see Fig. 2(g–j)). However, an increased degradation rate was observed at 2C charging (discharging at 1C in Fig. 2(h) and at 2C in Fig. 2(j)).

Summarising, the experimental works carried out with 122 cells allowed to obtain an extensive dataset which describes effectively the influence of temperature, DOD, and middle-SOC for a relatively broad operating window of Li-ion cells, which overlaps the typical operating conditions in many real applications. It is noteworthy that high C-rate levels, as well as negative temperatures are not represented in the data, which could limit the applicability of the developed model in such operating conditions (see Section 8). Furthermore, the additional tests realised at dynamic operating conditions allow to validate the performances of the model under time-varying stress-factors profiles, which are closer to real-world operation.

## 4. Gaussian process theory

This section aims to provide a brief overview of Gaussian Process models, introducing the main concepts and the predictive equations. Detailed explanations are available in [32].

The GP is a random process, *i.e.* a random entity whose realisation is a function $f(\mathbf{x})$ instead of a single value. Rather than assuming a parametric form for the function to fit the data, $f(\mathbf{x})$ is assumed to be a sample of a Gaussian random process distribution. Since the GP is a nonparametric model, even when observations have been added, the model is always able to fit the new upcoming data.

A GP is fully determined by its mean and covariance functions. Defining the mean function $m(\mathbf{x})$ and the covariance function $\kappa(\mathbf{x}, \mathbf{x}')$ of

## DOD dependency

## Middle-SOC dependency

## Charging C-rate dependency

## Discharging C-rate dependency

## Symmetric charging and discharging C-rate

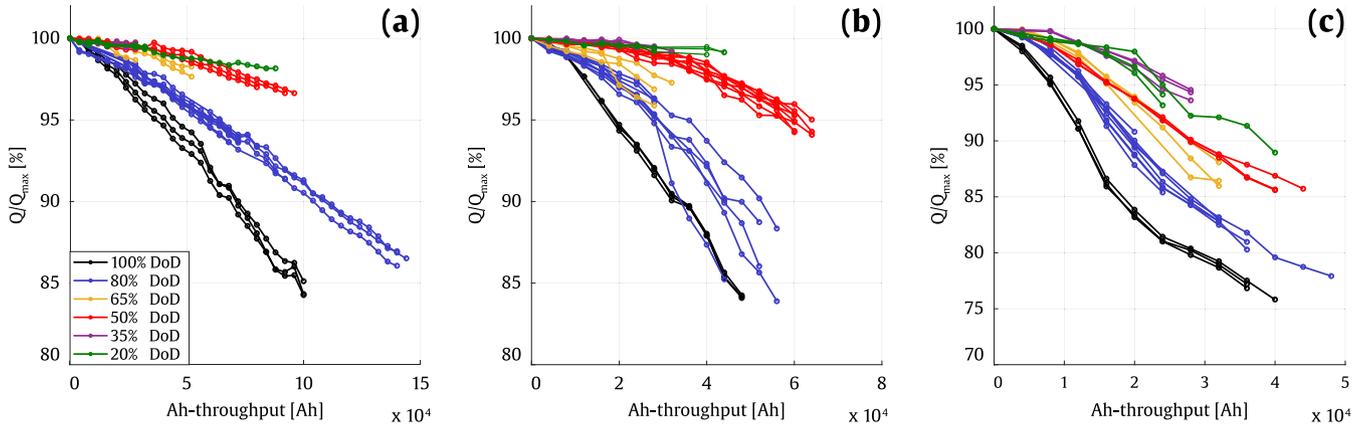
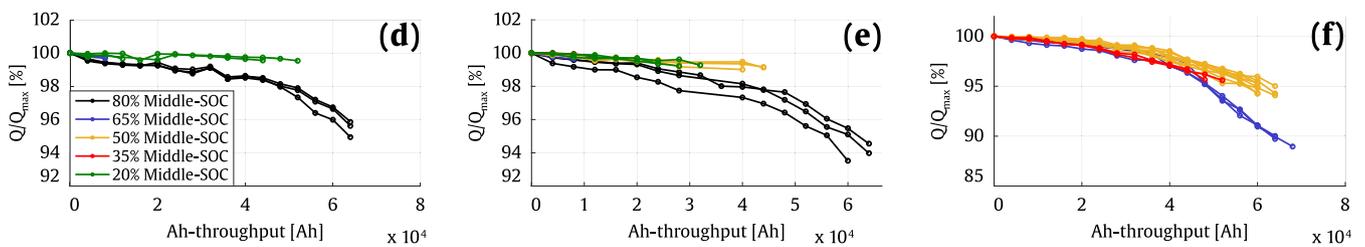
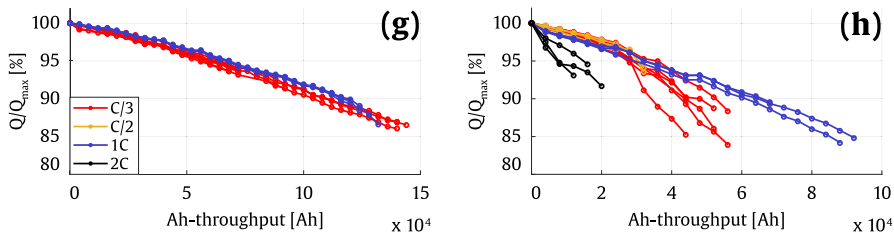
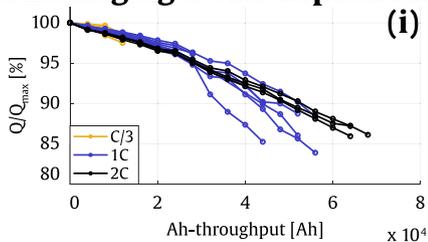
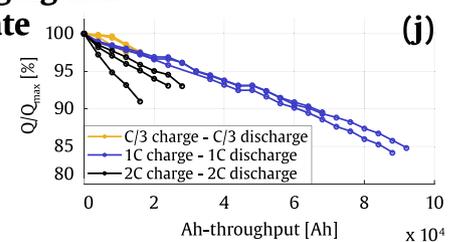

**Fig. 2.** Normalised capacity (with maximum value $Q_{max}$), after the preprocessing of the raw data obtained from the static ageing tests at (a) 25°C, 50% middle-SOC, C/3 – 1C, and several DOD values, (b) 35°C, 50% middle-SOC, C/3 – 1C, and several DOD values, (c) 45°C, 50% middle-SOC, C/3 – 1C, and several DOD values, (d) 35°C, 10% DOD, C/3 – 1C, and several middle-SOC values, (e) 35°C, 20% DOD, C/3 – 1C, and several middle-SOC values, (f) 35°C, 50% DOD, C/3 – 1C, and several middle-SOC values, (g) 25°C, 80% DOD, 50% middle-SOC, 1C discharging rate, and several charging rate values, (h) 35°C, 80% DOD, 50% middle-SOC, 1C discharging rate, and several charging rate values, (i) 35°C, 80% DOD, 50% middle-SOC, C/3 charging rate, and several discharging rate values, and (j) 35°C, 80% DOD, 50% middle-SOC and several symmetric charging and discharging rate values.

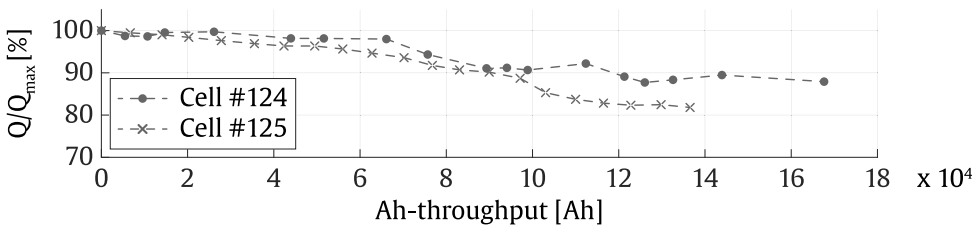

**Fig. 3.** Normalised capacity (with initial value $Q_{max}$), after the preprocessing of the raw data obtained from the dynamic ageing tests, for the cell #124 and #125.

a real process *f*(**x**) as:

$$m(\mathbf{x}) = \mathbb{E}[f(\mathbf{x})]$$
$$\kappa(\mathbf{x}, \mathbf{x}') = \mathbb{E}[(\mathbf{f}(\mathbf{x}) - \mathbf{m}(\mathbf{x}))(f(\mathbf{x}') - m(\mathbf{x}'))] \quad (1)$$

the GP can be expressed as

$$\mathbf{f}(\mathbf{x}) \sim \mathcal{GP}(\mathbf{m}(\mathbf{x}), \kappa(\mathbf{x}, \mathbf{x}')) \quad (2)$$

where **x** and **x**′ are two different input vectors.

Both mean and covariance functions encode the prior assumptions about the function to be learnt. They also express the expected behaviour of the model when the prediction inputs diverge from the inputs observed during training. The covariance function, also called the *kernel*, underpins the information about how relevant one target observation *y* of the training dataset is to predict the output *y*\*, on the basis of the similarity between their respective input values **x** and **x**\*.

The mean and covariance functions depend on some hyperparameters *θ*, which must be learnt from the training dataset. From a GP point of view, the mean and covariance function selection and learning the corresponding hyperparameters are the main tasks which must be carried out during the training phase. Hyperparameters are typically estimated by the maximisation of the marginal likelihood logarithm, using the gradient of the marginal likelihood with respect to such hyperparameters [32]. The marginal likelihood is defined as the integral of the likelihood times the prior.

Under the GP framework, the prior is Gaussian $\mathbf{f}|X \sim \mathcal{N}(\mathbf{0}, K)$, and the likelihood is a factorised Gaussian $\mathbf{y}|\mathbf{f} \sim \mathcal{N}(\mathbf{f}, \sigma_n^2 I)$, where **f** is the vector of latent function values as $\mathbf{f} = (f(\mathbf{x}_1, ..., \mathbf{x}_n))^T$; *X* is the matrix of the training input values; $\mathcal{N}$ is the Gaussian (normal) distribution; *K* is the covariance matrix for the (noise free) **f** values; **y** is the vector of the training target values; $\sigma_n^2$ is the noise variance and *I* is the identity matrix.

The obtained log marginal likelihood is expressed in Eq. (3)

$$\log p(\mathbf{y}|X) = -\frac{1}{2}\mathbf{y}^T(K + \sigma_n^2 I)^{-1}\mathbf{y} - \frac{1}{2}\log |K + \sigma_n^2 I| - \frac{n}{2}\log 2\pi \quad (3)$$

The GP predictive equations are expressed in Eqs. (4)–(6).

$$\mathbf{f}_*|X, \mathbf{y}, X_* \sim \mathcal{N}(\bar{\mathbf{f}}_*, \text{cov}(\mathbf{f}_*)) \quad (4)$$

with

$$\bar{\mathbf{f}}_* = \mathbf{m}(X_*) + K(X_*, X)[K(X, X) + \sigma_n^2 I]^{-1}(\mathbf{y} - \mathbf{m}(X)) \quad (5)$$

$$\text{cov}(\mathbf{f}_*) = K(X_*, X_*) - K(X_*, X)[K(X, X) + \sigma_n^2 I]^{-1}K(X, X_*) \quad (6)$$

where $\mathbf{f}_*$, $\bar{\mathbf{f}}_*$, and $\text{cov}(\mathbf{f}_*)$ are the GP posterior prediction, its corresponding mean and its covariance, respectively; $X_*$ is the matrix of test inputs; $\mathbf{m}(X)$ and $\mathbf{m}(X_*)$ are the vectors of mean functions for the training and test inputs respectively; $K(X, X)$, $K(X_*, X_*)$, and $K(X, X_*)$ are the covariance matrices between training inputs, the test inputs, and training and test inputs, respectively.

## 5. Development of the cycle ageing model

### 5.1. Assumptions and input selection

As stated in Section 3, this paper focuses on the modelling of the progressive capacity loss corresponding to the second phase represented in Fig. 1. The development of the model was based on the following assumptions:

i) The predominant ageing mechanism involved in such phase is the formation of the SEI layer on the anode surface, which could be moderated, accelerated or expanded by the cycling conditions, characterised by the values of the different stress-factors mentioned in Sections 2 and 3.

ii) The capacity loss is strongly dependent on the interactions between the different stress-factors, as described in Section 3.

As explained in the first paper of the series, corresponding to the calendar ageing model, the influence of the stress-factors should be considered introducing the corresponding values directly as an input [4]. Therefore, the model proposed in this section considered six inputs:

- ΔAh-throughput: the number of Ah-throughput for which the ageing is predicted.
- $T^{-1}$: the reciprocal of the temperature corresponding to the cycled Ah-throughput (for alignment to the Arrhenius law).
- DOD: the DOD level corresponding to the cycled Ah-throughput.
- Middle-SOC: the average SOC corresponding to the cycled Ah-throughput.
- Charging C-rate: the charging C-rate corresponding to the cycled Ah-throughput.
- Discharging C-rate: the discharging C-rate corresponding to the cycled Ah-throughput.

The output of the model was the capacity loss ΔQ corresponding to the ΔAh-throughput cycled at $T^{-1}$, DOD, Middle-SOC, Charging C-rate and Discharging C-rate conditions.

### 5.2. Kernel construction

As justified in [4], the framework of compositional kernels is a suitable solution to develop covariance functions tailored to Li-ion battery ageing application: a main kernel could be constructed composed of interpretable components, each one related to a specific input dimension [33]. In order to focus on the behaviour of the composed kernels, a zero-mean function was defined in this work. This is not a significant limitation, since the mean of the posterior process is not confined to be zero [32].

#### 5.2.1. Selecting individual kernel components

As explained in Section 4, the GP framework is a nonparametric model, and therefore the *learning* problem is the problem of finding the suitable properties of the function (isotropy, anisotropy, smoothness, etc.), rather than a particular functional form [32].

The range of the DOD and Middle-SOC input dimensions is intrinsically limited between 0–100%. Furthermore, the operation window corresponding to the Temperature, Charging C-rate and Discharging C-rate inputs is also limited by the recommendations of the manufacturer (e.g. cycling and storage temperatures between -30°C and 55°C), specified in the first paper of the series [4]. This is defined to be a local modelling problem and therefore the kernel components corresponding to the stress-factors' input spaces could be represented by *isotropic* kernels, as justified in [4]. Among the different isotropic kernels, the 5/2 Matérn kernels imply a suitable smoothness assumption to represent the physical processes inside the battery (as suggested in [4]), and were then selected to host independently the input dimensions corresponding to each stress-factor.

The kernel component related to the $\Delta Ah - throughput$ input dimension requires several $\Delta Ah - throughput$ values to be involved in the training data, in order to optimise the associated hyperparameter. In order to limit the training computation time, only three different values of $\Delta Ah - throughput$ were processed in the training data (which are 4000, 8000 and 12000 ΔAh). Table 2 illustrates the structure of the training data. In this context, the use of an isotropic kernel requires a large amount of different values of $\Delta Ah - throughput$ for long-term prediction, implying a large quantity of training data and increased computation times. Therefore, this kernel component should be anisotropic. In the second phase of the Li-ion cells ageing described in Fig. 1, the capacity loss seems to be linear with respect to $\Delta Ah - throughput$. Therefore, a linear kernel component was selected for this input dimension.

Although the data vectors 'CELL002 – data vector 1' and 'CELL002 – data vector 4' in Table 2 have the same inputs values, the target is different because both correspond to the capacity loss from a different starting point, in the capacity curve of the CELL002. The data vectors with identical input values and different outputs are useful for the determination of the noise hyperparameter of the GP models (see Eq. (7)).

Table 2
Example of the training data structure, relating the input data to the corresponding target.

| | | Input vector x | | | | | | Target y |
|---|---|---|---|---|---|---|---|---|
| | | $\Delta Ah - throughput$ [Ah] | $T^{-1}$ [$K^{-1}$] | DOD [%] | Middle-SOC [%] | Charging C-rate [C] | Discharging C-rate [C] | $\Delta Q$ [%] |
| CELL002 | Data vector 1 | 4000 | 0.0034 | 100 | 50 | C/3 | 1C | -0.163 |
| | Data vector 2 | 8000 | | | | | | -0.743 |
| | Data vector 3 | 12000 | | | | | | -1.101 |
| | Data vector 4 | 4000 | | | | | | -0.579 |
| | Data vector 5 | 8000 | | | | | | -0.937 |
| … | … | … | … | … | … | … | … | … |
| CELL055 | Data vector 1 | 4000 | 0.0032 | 50 | 35 | C/3 | 1C | -0.135 |
| | Data vector 2 | 8000 | | | | | | -0.142 |
| | Data vector 3 | 12000 | | | | | | -0.451 |
| | Data vector 4 | 4000 | | | | | | -0.007 |
| | Data vector 5 | 8000 | | | | | | -0.316 |
| | … | … | … | … | … | | | … |

*5.2.2. Composing the whole kernel*

In the GP framework, the kernel function is also a covariance function and therefore must be positive semidefinite [32]. Moreover, positive semidefinite compositional kernels are closed under the addition and multiplication of basic kernels. Additive kernels assume the added stochastic processes to be independent [33]. However, as specified in Section 5.1, the different inputs were assumed to have a strong interaction on their influence on the capacity loss, and hence an additive kernel composition should be avoided. In order to account for the interactions between the different input dimensions, the tensor product is suggested within [32,33] and is used in the composed kernel (Eq. (7)).

$$\kappa(\mathbf{x}, \mathbf{x}') = \sigma_f^2 \cdot \begin{bmatrix} \prod_{n=1}^{5} \left(1 + \sqrt{5} \cdot \frac{|x_n - x'_n|}{\theta_n} + \frac{5}{3} \cdot \frac{|x_n - x'_n|^2}{\theta_n^2}\right) \cdot \\ \exp\left(-\sqrt{5} \cdot \frac{|x_n - x'_n|}{\theta_n}\right) \\ \cdot (x_6 \cdot x'_6 + \theta_6^2) \end{bmatrix} + \sigma_n^2 \cdot I$$

(7)

where **x** and **x**' are different input vectors structured as $\mathbf{x} = (x_1, x_2, x_3, x_4, x_5, x_6)$, with $x_1 = T^{-1}$, $x_2 = DOD$, $x_3 = Middle - SOC$; $x_4 = Charging\ C - rate$; $x_5 = Discharging\ C - rate$ and $x_6 = \Delta Ah - throughput$.

$\theta_1$, $\theta_2$, $\theta_3$, $\theta_4$, $\theta_5$ and $\theta_6$ are the hyperparameters related to the corresponding input spaces. The additional hyperparameters $\sigma_f^2$ and $\sigma_n^2$ are respectively the signal variance, which plays the role of scaling the outputs in the dimension of the capacity loss $\Delta Q$, and the noise variance, which models an additive Gaussian noise from the data.

## 6. Learning from static operating conditions

This section aims to illustrate the ability of the developed GP model to improve its prediction performances while observing an increasing number of cycling data. Indeed, as new observations of cycling conditions are presented to the model, the training dataset of the model involves a more comprehensive view of the influence of the different combinations of stress-factors on the capacity loss. Therefore, for each prediction, the covariance function is able to find more similar examples in the stored training dataset, in term of cycling conditions. The prediction performances of the model improve throughout the whole operation window of the Li-ion cells.

In this section, the improvement of the model performances was evaluated in terms of:

i) **Accuracy of the prediction**: as the training dataset increases, a reduction of the prediction errors is expected over the whole operation window. The metrics used to evaluate the prediction error are detailed in Section 6.1.
ii) **Confidence in the prediction**: as the training dataset increases, the model disposes of more information about the ageing throughout the whole operation window. In accordance with the covariance Eq. (6), the confidence intervals of the predictions are expected to reduce, signifying that the model is more confident about its predictions. The metric used to evaluate the accuracy of the confidence intervals is detailed in Section 6.1.

*6.1. Evaluation metrics*

Six different metrics were used to assess the prediction performances of the two ageing models. The first one was the root-mean-square error (RMSE) of the output of the model, which was the capacity loss $\Delta Q$, defined according to Eq. (8).

$$RMSE_{\Delta Q}(\hat{y}_i, y_i) = \sqrt{\frac{1}{N_T} \sum_{i=1}^{N_T} (\hat{y}_i - y_i)^2}$$

(8)

where $\hat{y}_i$ is the predicted output, $y_i$ is the measured output and $N_T$ is the number of points to be evaluated. The second metric was defined as the RMSE of the predicted capacity curve:

$$RMSE_Q(\hat{Q}_i, Q_i) = \sqrt{\frac{1}{N_T} \sum_{i=1}^{N_T} (\hat{Q}_i - Q_i)^2}$$

(9)

where $\hat{Q}_i$ is the predicted capacity calculated by accumulation of the output and $Q_i$ is the measured capacity. This second metric is useful in order to evaluate the accumulative error of the model.

The RMSE is useful to assess the prediction performances of a model, with an emphasis on the high deviations which are strongly penalised. In order to evaluate the ability of the model to capture the main trends of the data, the analysis was completed with the implementation of the mean-absolute-error (MAE), defined in Eqs. (10) and (11) in terms of model output and capacity curve, respectively.

$$MAE_{\Delta Q}(\hat{y}_i, y_i) = \frac{1}{N_T} \sum_{i=1}^{N_T} |\hat{y}_i - y_i|$$

(10)

$$MAE_Q(\hat{Q}_i, Q_i) = \frac{1}{N_T} \sum_{i=1}^{N_T} |\hat{Q}_i - Q_i|$$

(11)

In the context of this study, the main objective of the model was to capture the main trends of Li-ion battery ageing in different operating conditions, rather than achieving a perfect fit of each data point.

Therefore, a 2% $MAE_Q$ threshold was defined as acceptable prediction error.

The final metric was the calibration score, which aimed at quantifying the accuracy of the uncertainty estimates. It is defined as the percentage of measured results in the test dataset that are within a predicted credible interval. Within a $\pm 2\sigma$ interval, corresponding to a 95.4% probability for a Gaussian distribution, the calibration score is given by Eqs. (12) and (13).

$$CS_{2\sigma-\Delta Q} = \frac{1}{N_T} \sum_{i=1}^{N_T} [|\hat{y}_i - y_i| < 2\sigma] \cdot 100 \quad (12)$$

$$CS_{2\sigma-Q} = \frac{1}{N_T} \sum_{i=1}^{N_T} [|\hat{Q}_i - Q_i| < 2\sigma] \cdot 100 \quad (13)$$

Therefore, $CS_{2\sigma}$ should be approximately 95.4% if the uncertainty predictions are accurate. Higher or lower scores indicate under- or over-confidence, respectively [8].

### 6.2. Training case studies to illustrate the learning of new operating conditions

Following the method introduced in the first paper of the series, 16 training cases were defined in order to illustrate how the GP model could learn from new observations and improve prediction performances. Each training case involved a different number of training data from the ageing dataset presented in Section 3. From the training case 1 to the training case 16, the number of training data increased: the data corresponding to new cycling conditions was included progressively, revealing one by one the influence of the different levels of the different stress-factors.

The distinct temperature values were introduced from case 1 to case 2, followed by the DOD levels from case 3 to case 7, the middle-SOC levels from case 8 to case 11, the charging C-rate levels from case 12 to case 14 and finally the discharging C-rate levels from case 15 to case 16. The introduction of each stress-factors level was guided by the following process: the highest level was introduced first, followed by the lowest level, and then the range was completed adjoining one by one the levels equidistant to the already known values, alternating the highest and lowest values. Illustrating the process in the DOD range: i) 100% DOD, the highest value, was already included in cases 1 and 2, then ii) the lowest value i.e. 20% DOD was included in case 3, iii) the equidistant would be 60% DOD, then the closest available values 65% and 50% DOD were included in case 4 and case 5 respectively, and iv) the highest (%80 DOD) and lowest (35% DOD) remaining levels were respectively added in cases 6 and 7. Notice that the 10% DOD level was included later, because the 50% middle-SOC level was not available at such DOD.

Table 3 indicates the characteristics of each training case. The different cells and the related cycling conditions involved during the training process are specified, as well as the corresponding ratio of the amount of training data with respect to the whole available data.

### 6.3. Prediction results

#### 6.3.1. Accuracy improvement

The black curves in Fig. 4 indicate the prediction accuracy of the GP model proposed in Section 5, trained with the different training cases defined in Section 6.2, in term of $MAE_{\Delta Q}$ and $MAE_Q$. The corresponding RMSE values are indicated in Table C1, Appendix C. For each training case, the error calculation was performed separately for:

i) The training cells: the mean value of the prediction errors obtained for all the cells involved in the training case was calculated (Fig. 4(a)). Such errors are informative about the ability of the model to fit the training data.

ii) The validation cells: the mean value of the prediction errors obtained for all the cells not involved in the training case was calculated (Fig. 4(b)). Such error is relevant to evaluate the generalisation ability of the model.

iii) Some targeted validation cells: the mean value of the prediction errors obtained for the validation cells which operated at unobserved levels of the partially explored stress-factors (Fig. 4(c)). For instance, the influence of the DOD is learned from the training case 3 to 7; in the training case 4, the training data included the data corresponding to the 20%, 65% and 100% DOD operation. Then the prediction error corresponding to the training case 4 plotted in Fig. 4(c) was calculated only for the validation cells corresponding to the 50%, 80% and 35% DOD cycling conditions, neglecting the errors corresponding to the cells cycled at different values of the further stress-factors. Such error is relevant to evaluate the generalisation ability of the model, to the extent of the partially explored input spaces.

iv) All the cells: the mean value of the prediction errors obtained for all the cells (Fig. 4(d)). Such error is informative about the global accuracy of the model.

As expected, the predictions errors of the training cells in Fig. 4(a) fulfil the 2% $MAE_Q$ threshold for all the training cases. Regarding the validation cells, the threshold of the 2% $MAE_Q$ is reached for the training case 4 (see Fig. 4(b)), and the performances of the model seem not to improve significantly since such training case.

Fig. 4(c) describes the evolution of the generalisation ability of the model throughout the whole range of each stress-factor. Focussing on the part related to the learning of the influence of the DOD, the first points correspond to the mean value of the MAE errors obtained with the GP model trained with training case 2 and performing predictions for all the cells tested at the cycling conditions corresponding to the learning of the DOD in Table 3. At this training stage, the model only observed the influence of cycling at 100% DOD, and then all the predictions at lower DOD values were overestimated, resulting in a high error of 4.89% $MAE_Q$. In the training case 3, the model started to learn the effect of the DOD by incorporating a 20% DOD condition in the training data. The mean error of the targeted validation cells improved drastically, as the model could infer from two different DOD values and gain a numerical intuition about the effect of the DOD on capacity loss. In the training case 4, the model possessed capacity loss values corresponding to 20%, 65% and 100% DODs in the training dataset. The mean error of the predictions corresponding to the cells at the remaining DOD values drop below the 2% $MAE_Q$ threshold, indicating a good generalisation of the model throughout the whole available range of DOD operation. Finally, the inclusion of new DOD values to the training dataset in the training cases 5 and 6 did not seem to significantly improve the generalisation ability of the model throughout the DOD operation range. Notice that for the training case 7, all the DOD values available from the dataset were involved in training, and therefore, there was no validation cells yet to evaluate the evolution of the generalisation ability of the model, and then the error cannot be calculated.

Regarding the evolution of the errors from training cases 7 to 10, which is related to the learning of the influence of the middle-SOC, the results were unaltered by the inclusion of new middle-SOC values in the training dataset (Fig. 4(c)). This is explainable by the relatively reduced influence on the capacity loss assigned by the model to the middle-SOC stress-factor (more details in Section 6.3.3). Furthermore, concerning the learning of the charging C-rate, an increase of the error is observable from training case 11 to 12, before the final reduction in case 13. This is due to the initial inclusion of the 2C charging condition in training case 12, which presents a faster capacity loss compared to the remaining levels of charging C-rate (see Fig. 2(h)). At this stage, the model tends to overestimate the ageing at intermediate charging C-rate values. This is corrected in the training case 13 by the incorporation of the 1C charging data.

**Table 3**
Summary of the different case studies, specifying the different cells involved and the related cycling conditions, as well as the ratio of the amount of training data with respect to the whole available data.

| | | Learning Temperature | Learning DOD | Learning MidSOC | Learning charging C-rate | Learning discharging C-rate | # Training data / # Total data [%] |
|---|---|---|---|---|---|---|---|
| **CASE 1** | T | 25  45 | | | | | 7.56 |
| | DOD | 100 | | | | | |
| | MidSOC | 50 | | | | | |
| | C-rate CHA | C/3 | | | | | |
| | C-rate DCH | 1C | | | | | |
| **CASE 2** | T | 25  45  35 | | | | | 10.08 |
| | DOD | 100 | | | | | |
| | MidSOC | 50 | | | | | |
| | C-rate CHA | C/3 | | | | | |
| | C-rate DCH | 1C | | | | | |
| **CASE 3** | T | 25  45  35 | 25, 35, 45 | | | | 15.70 |
| | DOD | 100 | 20 | | | | |
| | MidSOC | 50 | 50 | | | | |
| | C-rate CHA | C/3 | C/3 | | | | |
| | C-rate DCH | 1C | 1C | | | | |
| **CASE 4** | T | 25  45  35 | 25, 35, 45 | | | | 21.13 |
| | DOD | 100 | 20  65 | | | | |
| | MidSOC | 50 | 50 | | | | |
| | C-rate CHA | C/3 | C/3 | | | | |
| | C-rate DCH | 1C | 1C | | | | |
| **CASE 5** | T | 25  45  35 | 25, 35, 45 | | | | 37.22 |
| | DOD | 100 | 20  65  50 | | | | |
| | MidSOC | 50 | 50 | | | | |
| | C-rate CHA | C/3 | C/3 | | | | |
| | C-rate DCH | 1C | 1C | | | | |
| **CASE 6** | T | 25  45  35 | 25, 35, 45 | | | | 59.97 |
| | DOD | 100 | 20  65  50  80 | | | | |
| | MidSOC | 50 | 50 | | | | |
| | C-rate CHA | C/3 | C/3 | | | | |
| | C-rate DCH | 1C | 1C | | | | |
| **CASE 7** | T | 25  45  35 | 25, 35, 45 | | | | 64.25 |
| | DOD | 100 | 20  65  50  80  35 | | | | |
| | MidSOC | 50 | 50 | | | | |
| | C-rate CHA | C/3 | C/3 | | | | |
| | C-rate DCH | 1C | 1C | | | | |
| **CASE 8** | T | 25  45  35 | 25, 35, 45 | 35 | | | 71.05 |
| | DOD | 100 | 20  65  50  80  35 | 10, 20 | | | |
| | MidSOC | 50 | 50 | 80 | | | |
| | C-rate CHA | C/3 | C/3 | C/3 | | | |
| | C-rate DCH | 1C | 1C | 1C | | | |
| **CASE 9** | T | 25  45  35 | 25, 35, 45 | 35 | | | 74.03 |
| | DOD | 100 | 20  65  50  80  35 | 10, 20  10, 20 | | | |
| | MidSOC | 50 | 50 | 80  20 | | | |
| | C-rate CHA | C/3 | C/3 | C/3 | | | |
| | C-rate DCH | 1C | 1C | 1C | | | |
| **CASE 10** | T | 25  45  35 | 25, 35, 45 | 35 | | | 78.00 |
| | DOD | 100 | 20  65  50  80  35 | 10, 20  10, 20  10, 20, 50 | | | |
| | MidSOC | 50 | 50 | 80  20  65 | | | |
| | C-rate CHA | C/3 | C/3 | C/3 | | | |
| | C-rate DCH | 1C | 1C | 1C | | | |
| **CASE 11** | T | 25  45  35 | 25, 35, 45 | 35 | | | 80.11 |
| | DOD | 100 | 20  65  50  80  35 | 10, 20  10, 20  10, 20, 50  20, 50 | | | |
| | MidSOC | 50 | 50 | 80  20  65  35 | | | |
| | C-rate CHA | C/3 | C/3 | C/3 | | | |
| | C-rate DCH | 1C | 1C | 1C | | | |
| **CASE 12** | T | 25  45  35 | 25, 35, 45 | 35 | 35 | | 80.88 |
| | DOD | 100 | 20  65  50  80  35 | 10, 20  10, 20  10, 20, 50  20, 50 | 80 | | |
| | MidSOC | 50 | 50 | 80  20  65  35 | 50 | | |
| | C-rate CHA | C/3 | C/3 | C/3 | 2C | | |
| | C-rate DCH | 1C | 1C | 1C | 1C | | |
| **CASE 13** | T | 25  45  35 | 25, 35, 45 | 35 | 35  25, 35 | | 92.41 |
| | DOD | 100 | 20  65  50  80  35 | 10, 20  10, 20  10, 20, 50  20, 50 | 80 | | |
| | MidSOC | 50 | 50 | 80  20  65  35 | 50 | | |
| | C-rate CHA | C/3 | C/3 | C/3 | 2C  1C | | |
| | C-rate DCH | 1C | 1C | 1C | 1C  1C | | |
| **CASE 14** | T | 25  45  35 | 25, 35, 45 | 35 | 35  25, 35  35 | | 94.09 |
| | DOD | 100 | 20  65  50  80  35 | 10, 20  10, 20  10, 20, 50  20, 50 | 80 | | |
| | MidSOC | 50 | 50 | 80  20  65  35 | 50 | | |
| | C-rate CHA | C/3 | C/3 | C/3 | 2C  1C  C/2 | | |
| | C-rate DCH | 1C | 1C | 1C | 1C  1C | | |
| **CASE 15** | T | 25  45  35 | 25, 35, 45 | 35 | 35  25, 35  35 | 35 | 98.83 |
| | DOD | 100 | 20  65  50  80  35 | 10, 20  10, 20  10, 20, 50  20, 50 | 80 | 80 | |
| | MidSOC | 50 | 50 | 80  20  65  35 | 50 | 50 | |
| | C-rate CHA | C/3 | C/3 | C/3 | 2C  1C  C/2 | C/3  2C | |
| | C-rate DCH | 1C | 1C | 1C | 1C  1C | 2C  2C | |
| **CASE 16** | T | 25  45  35 | 25, 35, 45 | 35 | 35  25, 35  35 | 35 | 100.00 |
| | DOD | 100 | 20  65  50  80  35 | 10, 20  10, 20  10, 20, 50  20, 50 | 80 | 80 | |
| | MidSOC | 50 | 50 | 80  20  65  35 | 50 | 50 | |
| | C-rate CHA | C/3 | C/3 | C/3 | 2C  1C  C/2 | C/3  2C  C/3 | |
| | C-rate DCH | 1C | 1C | 1C | 1C  1C | 2C  2C  C/3 | |

Fig. 5(a–e) illustrate the capacity loss predictions of the GP model resulting from the training case 4, for different cycling conditions involved in the training data. The average $MAE_{\Delta Q}$ and $MAE_Q$ errors of the model corresponding to the training case 4 were 0.68% and 1.12%, respectively, for the training cells. The average $CS_{2\sigma-\Delta Q}$ and $CS_{2\sigma-Q}$ were respectively 90.65% and 77.75%. Furthermore, Fig. 5(f–j) depict the capacity loss predictions of the GP model resulting from the training case 4, for different validation cycling conditions, which were not involved in the training data. The average $MAE_{\Delta Q}$ and $MAE_Q$ errors of the model corresponding to the training case 4 were 0.55% and 1.02%, respectively, for the validation cells. The average $CS_{2\sigma-\Delta Q}$ and $CS_{2\sigma-Q}$ were respectively 94.02% and 82.01%. Fig. 5(k–o) aims to underpin the improvement of the generalisation performances of the GP, while

# Learning the influence of

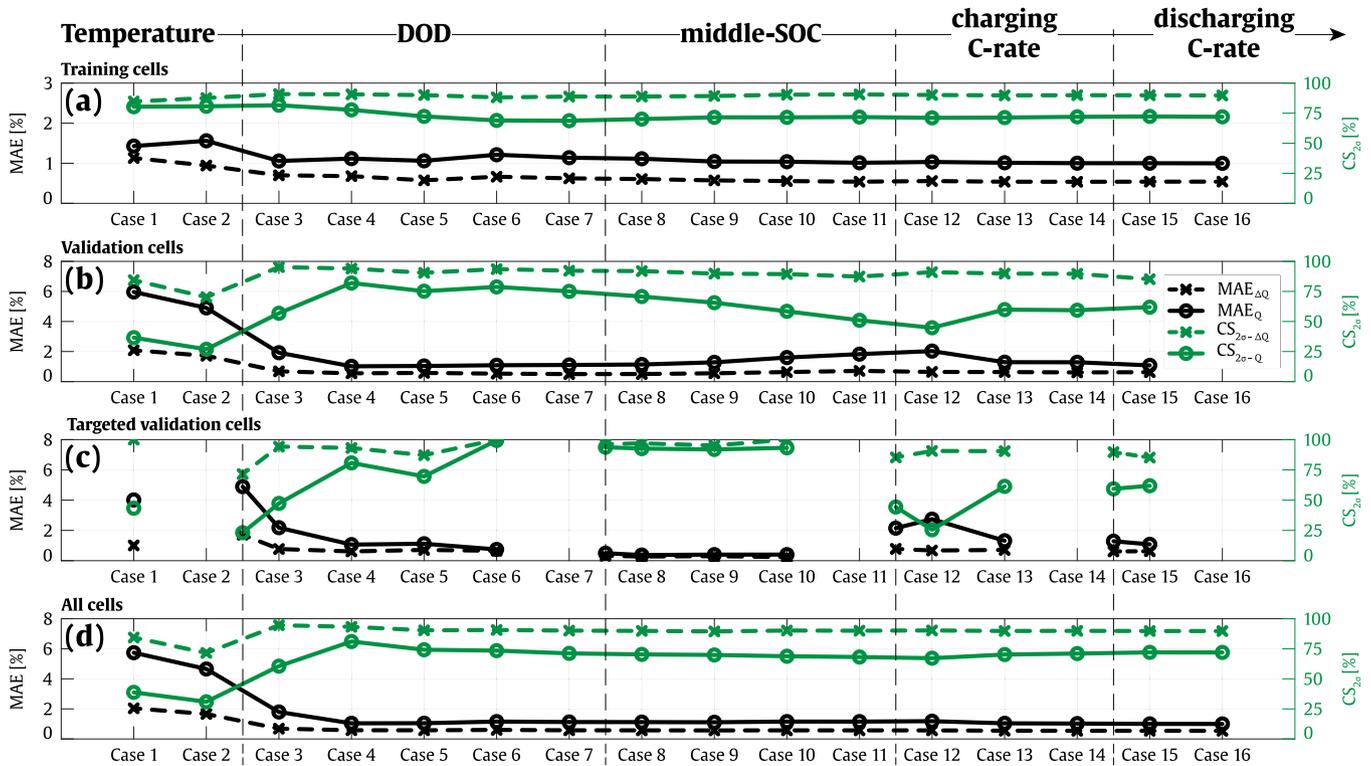

**Fig. 4.** Prediction results corresponding to each training case, in term of MAE and $CS_{2\sigma}$, distinguishing the errors of (a) all the training cells, (b) all the validation cells, (c) targeted validation cells and (d) all the cells.

increasing the number of training values in the input space of the DOD. To this end, the capacity loss predictions were represented for the cells #004 to #011 (which operated at 25°C, 80% DOD, 50% middle-SOC and C/3 – 1C charging and discharging C-rates), using GP models obtained from different training cases.

As previously explained, the models obtained from the training cases 1 and 2 did not have any information about the effect of the DOD on the capacity loss, as the training data involved the single input of 100% DOD. At this stage, the prediction at lower DOD levels were overestimated (see Fig. 5(k) and (l)), The mean error in such condition was 3.91% and 3.88% $MAE_Q$, respectively. In the training cases 3 and 4, the integration of the 20% and 65% DOD operating conditions in the training dataset allowed improving the predictions at 80% DOD, reaching 2.34% and 0.42% $MAE_Q$ values, respectively (see Fig. 5(m) and (n)). For comparison, the results obtained with a fully trained GP (training case 7) were also plotted in Fig. 5(o): there was not significant improvement in term of error reduction. However, the confidence intervals were slightly reduced, indicating a higher confidence of the model to perform predictions at 80% DOD, since such operating condition was represented in the training data (more details in Section 6.3.2). At this point, it is noteworthy that the model corresponding to the training case 7 is only used in this study for a sake of comparison with the previous cases. In fact, such a model would be unreliable for deployment, as all the available DOD levels were observed in training and then the generalisation ability of the model could not be validated in the space of DOD.

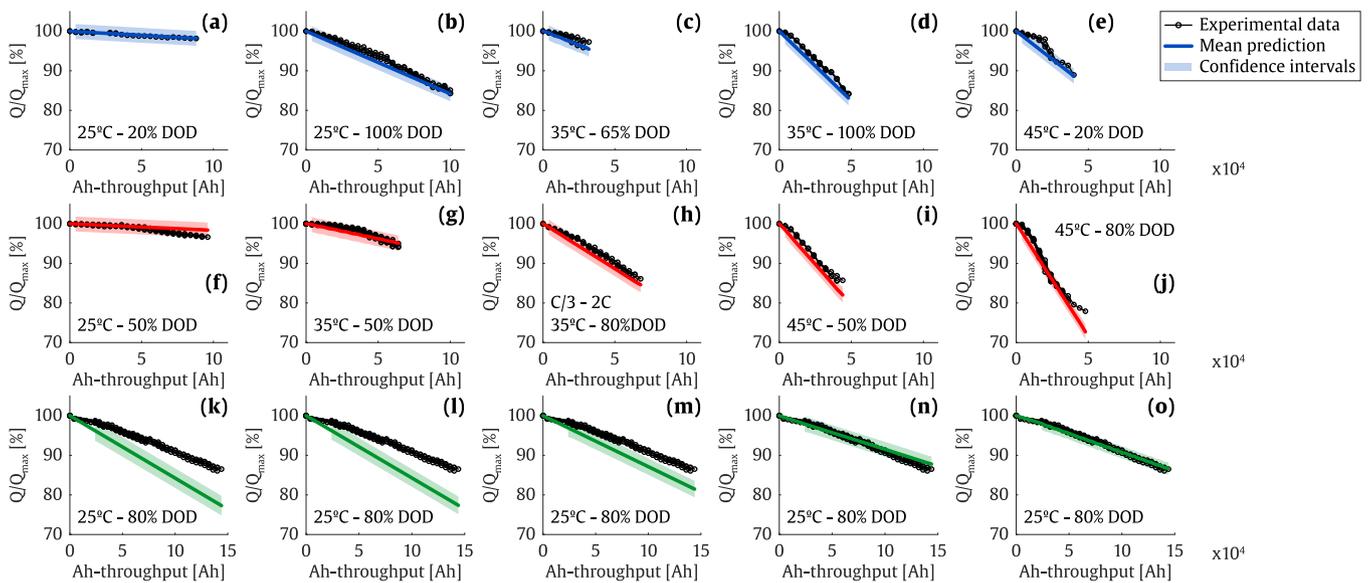

**Fig. 5.** (a–e) Capacity predictions with the GP model trained at training case 4, for the training cells cycled at the Temperature and DOD levels indicated in each graph. (f–j) Capacity predictions with the GP model trained at training case 4, for the validation cells cycled at the Temperature and DOD levels indicated in each graph. (k–o) Capacity predictions for the cells cycled at 25°C and 80% DOD, with the GP models trained at (k) training case 1, (l) training case 2, (m) training case 3, (n) training case 4 and (o) training case 7. Unless otherwise specified, the cells involved in (a–o) were cycled at 50% Middle-SOC, C/3 charging C-rate and 1C discharging C-rate.

### 6.3.2. Increase of confidence

According to the variance equation Eq. (6), the confidence intervals of a prediction reduce if the training dataset involves data samples similar to the predicted input values. Informally, this means that the model feels more confident to do predictions in case it already observed similar operating conditions in training data. Therefore, the analysis of the width of the confidence intervals – or equivalently the standard deviation value - along a large operating range of each stress-factor is informative about how confident the model feels to perform predictions throughout a broad operating window. In this sense, the evolution of the standard deviation throughout the input space testifies about the learning process of the model.

In Fig. 6, the evolution of the standard deviation of the GP model predictions is depicted throughout the whole operation window of the Li-ion cell under study, for the different training cases. For the model obtained from the training case 1, the standard deviation indicates lowest values around 25°C and 45°C, Fig. 6(a), which are the only temperatures experienced at this stage. The observation of the effect of a 35°C operation in the training case 2 flattened the curve around the such temperature: at this stage, the obtained model felt relatively confident to perform predictions within the 20°C–50°C temperature range. Notice that the model presented high standard deviation values at low and negative temperatures, due to the lack of information in such cycling regions. Fig. 6(b) corresponds to the learning of the influence of the DOD. As expected, the lowest standard deviation stood near 20% and 100% for training case 3, and the observation of intermediate DOD levels from the training cases 4 to 7 lead to reduced values in the whole range, unless below 20% DOD operation which still was an unknown cycling condition. Identical interpretation could be done from Fig. 6 (c–e) regarding the evolution of the standard deviation in the operation ranges of the middle-SOC, charging and discharging C-rate, respectively.

The reduction of the standard deviation in Fig. 6 testifies about the increment of the model's confidence to perform prediction throughout a broad operating window, as input spaces are progressively explored. Moreover, the accuracy of the confidence level of the model was evaluated using the calibration score metric, introduced in Section 6.1. As previously explained, the $CS_{2\sigma}$ values should be approximately 95.4% if the uncertainty predictions are accurate. Higher or lower scores indicate under- or over-confidence of the model, respectively [8].

In Fig. 4, the evolution of the mean value of the calibration scores are plotted for each training case of the GP model, in term of capacity loss and accumulated capacity. Since the training case 4, the overall $CS_{2\sigma-Q}$ values converge into approximately 75% (Fig. 4(d)). This traduces a slightly over-confident behaviour of the model in term of the accumulated capacity. However, regarding the calibration scores values corresponding to the output the model, the overall $CS_{2\sigma-\Delta Q}$ values converge into approximately 90%.

### 6.3.3. Sensitivity of the capacity loss to the stress-factors

Isotropic covariance functions implement automatic relevance

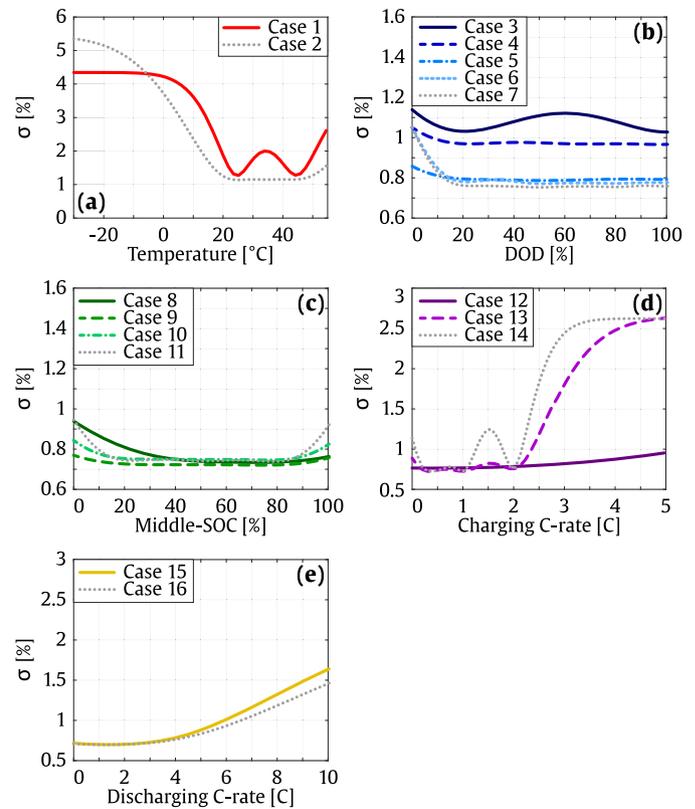

**Fig. 6.** Evolution of the standard deviations of the GP model predictions throughout the whole operation window of the Li-ion cell under study, from training case 1 to 16. (a) Evolution throughout the temperature space, at constant 80% DOD, 50% middle-SOC and C/3 – 1C charging and discharging C-rate (b) Evolution throughout the DOD space, at constant 35°C, 50% middle-SOC and C/3 – 1C charging and discharging C-rate (c) Evolution throughout the middle-SOC space, at constant 35°C, 20% DOD and C/3 – 1C charging and discharging C-rate (d) Evolution throughout the space of the charging C-rate, at constant 35°C, 80% DOD, 50% middle-SOC and 1C discharging C-rate and (e) Evolution throughout the space of the discharging C-rate, at constant 35°C, 80% DOD, 50% middle-SOC and C/3 charging C-rate.

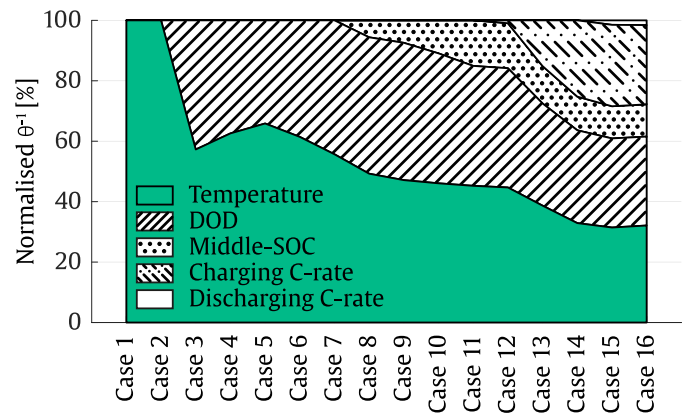

**Fig. 7.** Evolution of the relative relevance of the different stress-factors, from the training case 1 to 16.

determination, since the inverse of the length-scale determines how relevant an input is: if the length-scale has a very large value, the covariance will become almost independent of that input, effectively removing it from the inference [8]. Therefore, the sensitivity of the capacity loss to the different stress-factors could be analysed by observing the inverse of their respective hyperparameters. Fig. 7 displays, for each training case, the inverse of the hyperparameters

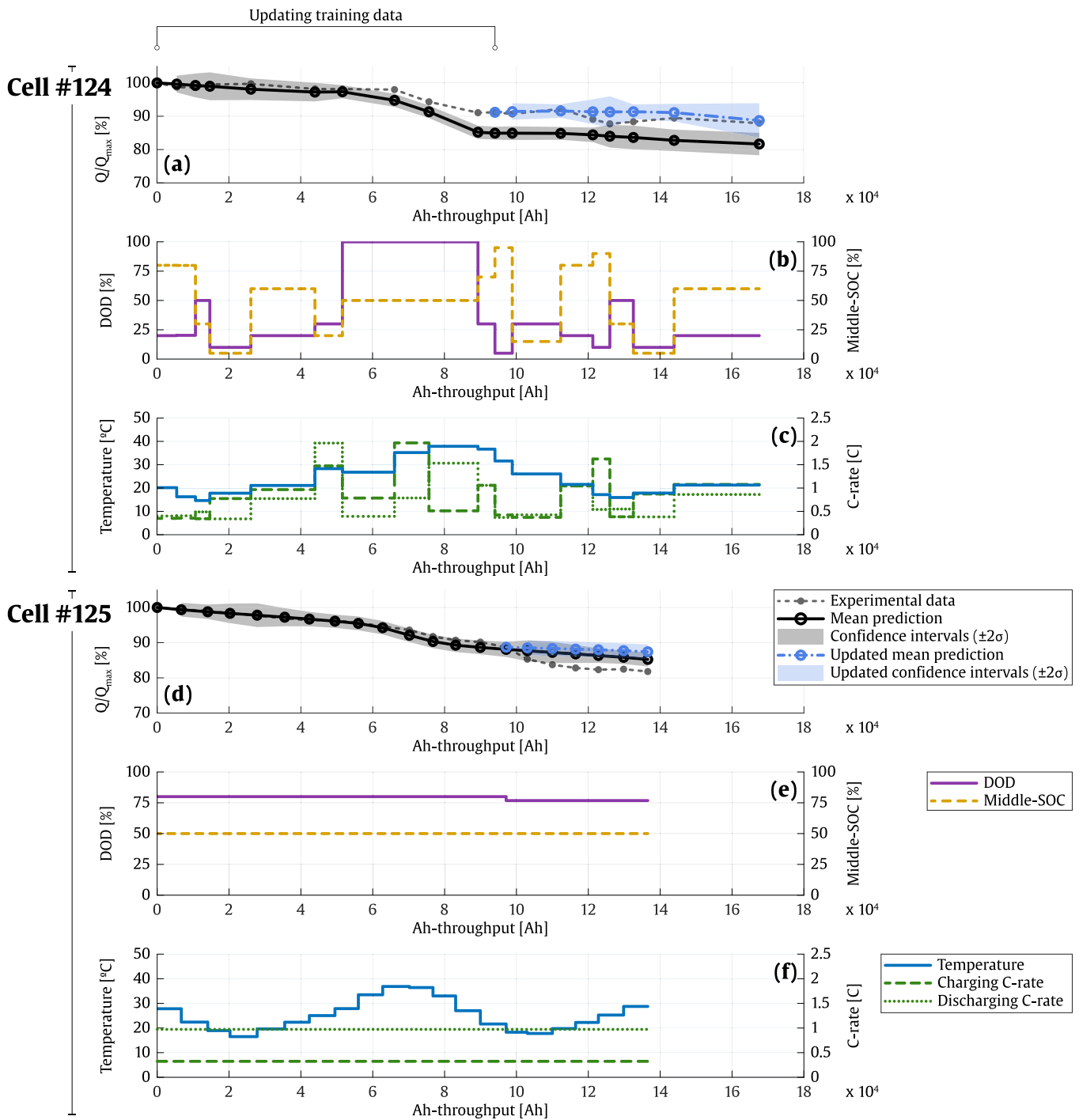

**Fig. 8.** (a) Normalised capacity (with maximum value $Q_{max}$) data and the corresponding ageing predictions for the initial model (training case 4, black line and grey area) and the updated model (blue line and area), for the cell #124. (b) DOD and middle-SOC profiles and (c) temperature and charging and discharging C-rate profiles applied to the cell #124. (d) Normalised capacity (with maximum value $Q_{max}$) data and the corresponding ageing predictions for the initial model (training case 4, black line and grey area) and the updated model (blue line and area), for the cell #125. (e) DOD and middle-SOC profiles and (f) temperature and charging and discharging C-rate profiles applied to the cell #125.(For interpretation of the references to colour in this figure legend, the reader is referred to the web version of this article.)

corresponding to each input dimension, relatively normalised to each other.

Fig. 7 illustrates the relative relevance of the different stress-factors, for the GP model corresponding to training case 1 to 16. In the training cases 1 and 2, only the temperature involved different operating values in the training dataset, as a single value was available for the remaining stress-factors. In absence of data to guide the optimisation of the corresponding hyperparameters, a high initial hyperparameter value was imposed to those stress-factors, in order to hinder their optimisation and then remove their effect from inference. In this context, the unique relevant stress-factor for the GP model was the temperature.

From the training case 3 to 7, different DOD levels were progressively included in the training dataset, and the corresponding hyperparameter was 'released' for optimisation. In Fig. 7, it could be observed that the relative relevance of the DOD input with respect to the capacity loss increased; however, the temperature variations was still considered slightly more impactful on the capacity loss than DOD variations. From the training case 8 to 11, the evolution of the importance assigned to the middle-SOC is observable, which was still limited compared to the temperature and DOD. In training case 12, a reduced impact of the charging rate was inferred, considering the difference in capacity losses between C/3 and 2C training data. However, the observation of the 1C and C/2 charging rates in training cases 13 and 14, which both lead to similar capacity loss as C/3 charging rate, suggested that all such difference stood between 1C and 2C: from this new perspective, small changes of charging C-rate induces relatively high changes in capacity loss, traducing a high covariance between these two variables. Then the GP assigned high relevance to the charging C-rate input in the training case 14. Finally, a reduced dependence of the capacity loss on the discharging C-rate was captured from the training cases 15 and 16, which is in accordance with the observations done in Section 3.

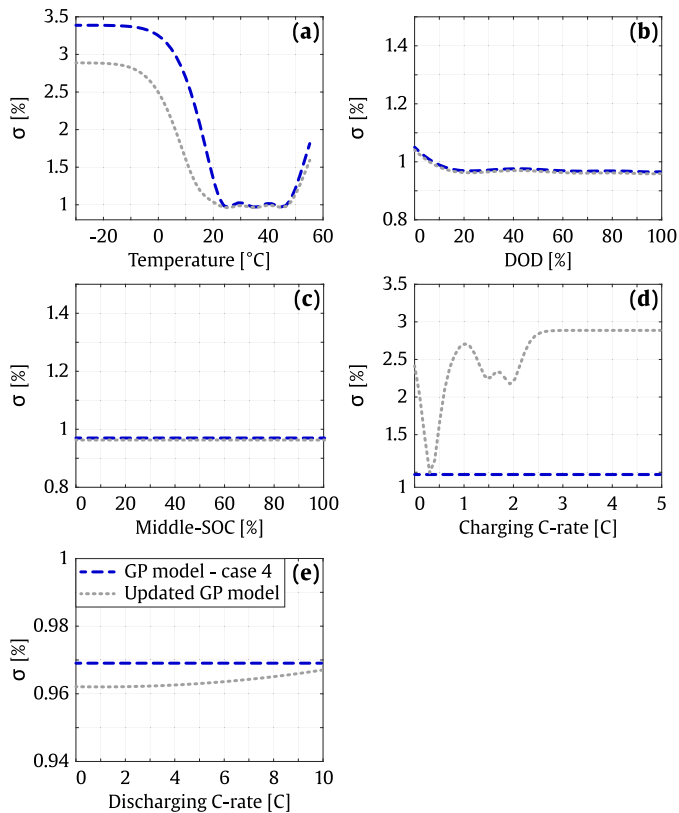

**Fig. 9.** Evolution of the standard deviations of the GP model predictions throughout the whole operation window of the Li-ion cell under study, from the model trained at case 4 to the model updated at dynamic operating conditions. (a) Evolution throughout the temperature space, at constant 80% DOD, 50% middle-SOC and C/3–1C charging and discharging C-rate (b) Evolution throughout the DOD space, at constant 35°C, 50% middle-SOC and C/3–1C charging and discharging C-rate (c) Evolution throughout the middle-SOC space, at constant 35°C, 20% DOD and C/3 – 1C charging and discharging C-rate (d) Evolution throughout the space of the charging C-rate, at constant 35°C, 80% DOD, 50% middle-SOC and 1C discharging C-rate and (e) Evolution throughout the space of the discharging C-rate, at constant 35°C, 80% DOD, 50% middle-SOC and C/3 charging C-rate.

In this way, the fully trained GP classified the relevance of the different stress-factors with respect to the capacity loss prediction in this order: 1/ temperature, 2/ DOD, 3/ charging C-rate, 4/ middle-SOC and 5/ discharging C-rate. At this point, it is important to highlight that although such comparison could clarify how the GP model understand the data, it does not imply causality.

## 7. Learning from dynamic operating conditions

As the operating conditions of Li-ion batteries are barely constant in real applications, the ageing models developed in the basis of ageing tests realised at constant operating conditions must be validated at dynamic operating conditions. Furthermore, as this study focuses on the development of ageing models oriented to learn from ageing data collected from real-world operation, the analysis of the possibility to infer about the correlations among the different stress-factors and the capacity loss directly from dynamic operation profile is necessary. To this end, the model developed in Section 5 was employed to perform ageing predictions for cells #124 and #125, the operating profiles of which were presented in Fig. 8(b, c) and (e, f), respectively. For the training case 4 (see Section 6), the GP model reached satisfying prediction results, achieving errors below the defined 2% $MAE_Q$ threshold. In this section, such training case was therefore selected as initial state of the model, in order to evaluate the prediction performances of the model at dynamic operating conditions. The obtained predictions are presented in black line (mean prediction) and grey area (confidence intervals) in Fig. 8(a) and (d), for the cells #124 and 125 respectively.

The model obtained from training case 4 achieved 1.13% and 0.46% errors in terms of $MAE_{\Delta Q}$, and 3.76% and 1.46% in terms of $MAE_Q$, for the cells #124 and #125 respectively. At approximately 90000 Ah-throughput of cycling, the whole range of the temperature profile was experienced for the cells #124 and #125. For the cell #124, different combinations of the remaining stress-factors were also observed, some of them reproduced on the remaining cycling profiles (e.g the combinations between ca. 11000-43000 Ah-throughput, were reproduced between ca. 126000-167000 Ah-throughput). Such point was then deemed to be a suitable updating point for the model, to be able to evaluate the learning ability of the model at dynamic operating conditions. Therefore, the operating conditions as well as the corresponding capacity loss values observed between 0–90000 Ah-throughput were included in the training dataset in order to obtain an updated GP model.

In Fig. 8(a) and (d), the blue curves represent the predictions performed with the updated model, for the cells #124 and #125 respectively. For the cell #125, only the temperature profile was varying, the remaining stress-factors beeing constant. The initial model predicted larger confidence intervals at cold temperatures (between 15°C–25°C), as the coldest temperature experienced in the training case 4 was 25°C. The observation of the such values increased the confidence of the model to perform predictions in this range. This is traduced in Fig. 8(d) by reduced confidence intervals at cold temperatures, compared with the initial predictions.

The cell #124 was cycled at dynamic temperatures, DOD, middle-SOC and charging and discharging C-rates profiles (see Fig. 8(b) and (c)). In Fig. 8(a), it could be observed that while the confidence intervals were reduced at some point (e.g. around 132000 Ah-throughput), they became larger at some other points (e.g. around 167000 Ah-throughput). In fact, in the training case 4 only different temperature and DOD values were observed, and the remaining stress-factors were then neglected from inference by imposing high initial hyperparameters (as explained in Section 6.3.3). When updating the model with the different stress-factors combinations observed in the dynamic profiles, all the stress-factors were involved in the learning process, and the confidence of the model for predicting throughout the whole operating window was modified. This is observable in Fig. 9, which reflects the evolution of the standard deviation of the model's predictions, for the model corresponding to the training case 4 and the model updated with the data obtained from dynamic operating profile until 90000 Ah-throughput. Regarding the range of the cycling temperatures, Fig. 9(a), it is remarkable that the model gained confidence around approximately 15°C–25°C, which is reflected by a reduction of the standard deviation in such region. Furthermore, a strong influence of the charging C-rate was detected from the dynamic profiles, leading to a large variability on the standard deviation even for small charging C-rate variations, **Fig. 9**(d). This explains why the confidence intervals became larger in some prediction point in **Fig. 8**(a): around 167000 Ah-throughput, for instance, the updated model predicted larger confidence intervals, because the ~1C charging C-rate value was identified as an 'uncertain' region for prediction due to i) the lack of training data in such charging rates region and ii) to the high influence of this stress-factor on the capacity loss, which was inferred from the previously observed ageing at dynamic operating profile.

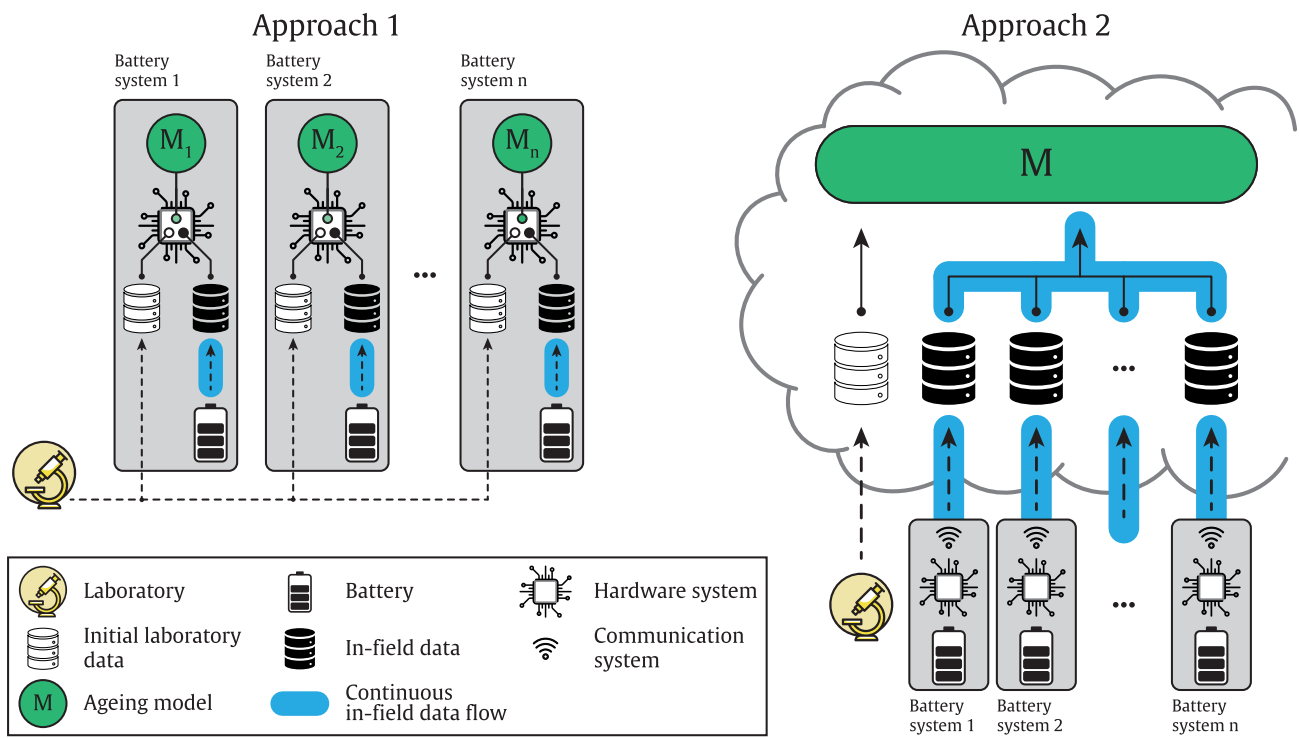

**Fig. 10.** Two different approaches for the deployment of ageing models in real applications. The first approach consists on the implementation of the ageing model in the local hardware device of each battery system. The second approach contemplates the communication and storage of the battery operation data to a data server in the cloud, in which the ageing model is implemented.

## 8. Discussion, limitation of the study and further works

The model developed in Section 5 demonstrated suitable performances to fit the data, independently from the number of training data and involved stress-factors. This is observable in **Fig. 4**(a), where both $MAE_{\Delta Q}$ and $MAE_Q$ curves of the training cells showed a constant level under the defined 2% threshold, from the training case 1 to 16.

The minimum amount of experimental ageing tests necessary from the laboratory for the development of the initial ageing model was determined: the training case 4 seems to present an adequate trade-off between the performances and the development cost of the model, insofar as the cell is used at the operating conditions recommended by the manufacturer (specified in the first paper of the series [4]). In fact, the $MAE_Q$ values dropped below the 2%, and the performances of the model seem not to improve significantly since such training case (see **Fig. 4**(b) and (d)). However, the operation at cold temperatures (between 0°C–25°C) was not contemplated in the experimental tests. The operating window of such initial ageing model should then be limited above circa 25°C, at least until the further learning of the influence of colder temperatures. Furthermore, the predictions of the model obtained from training case 4 are insensitive to the middle-SOC, charging and discharging C-rate variations. Although this does not seem to matter regarding the middle-SOC stress-factor, it could be problematic for the applications involving high charging and discharging C-rates. In such cases, supplementary laboratory tests could be necessary at several C-rate values.

The analysis of the uncertainty boundaries corroborates the findings observed in the first paper of the series: the reduction of the standard deviation in **Figs. 6** and **9** testified about the increment of the model's confidence to perform prediction throughout a broad operating window, as input spaces are progressively explored. Again, the developed GP model turned out to be slightly over-confident, according to the calibration scores curves represented in **Fig. 4**. As previously explained, the $CS_{2\sigma}$ values should be approximately 95.4% if the uncertainty predictions are accurate: the obtained $CS_{2\sigma-Q}$ and $CS_{2\sigma-\Delta Q}$ values converged approximately into 75% and 90% respectively (**Fig. 4**(d)). It could be observed that the confidence intervals of the model output are relatively close to the target value of 95.4%. The difference between the $CS_{2\sigma-Q}$ and $CS_{2\sigma-\Delta Q}$ suggests that the over-confidence of the model is induced by the error accumulation of the iterative prediction process. Therefore, further investigations would be required in order to study the propagation of model's uncertainty throughout the long-term ageing prediction [34].

In Section 6.3.3, the evolution of the hyperparameters' reciprocal was analysed, in order to illustrate how the model would actually be learning about the sensitivity of the capacity loss to each individual stress-factor. Nevertheless, it is noteworthy that, manipulating the Eq. (7), which corresponds to the developed covariance function, some terms involving the products among the different stress-factors' hyperparameters appear. Such terms could be interpreted as the covariance components corresponding to the interactions between the different stress-factors. The sensitivity analysis of the capacity loss to the stress-factors could be extended by involving such covariance components, in order to have a feedback about which combinations of stress-factors levels are most critical according to the GP model. Such analysis would be difficult to carry out with laboratory data, mainly due to the large amount of ageing data it would require. However, the incorporation of the real-world data collected from the deployed battery-packs could make such analysis possible. This could provide insightful inputs for the development of effective energy management strategies.

## DOD dependency

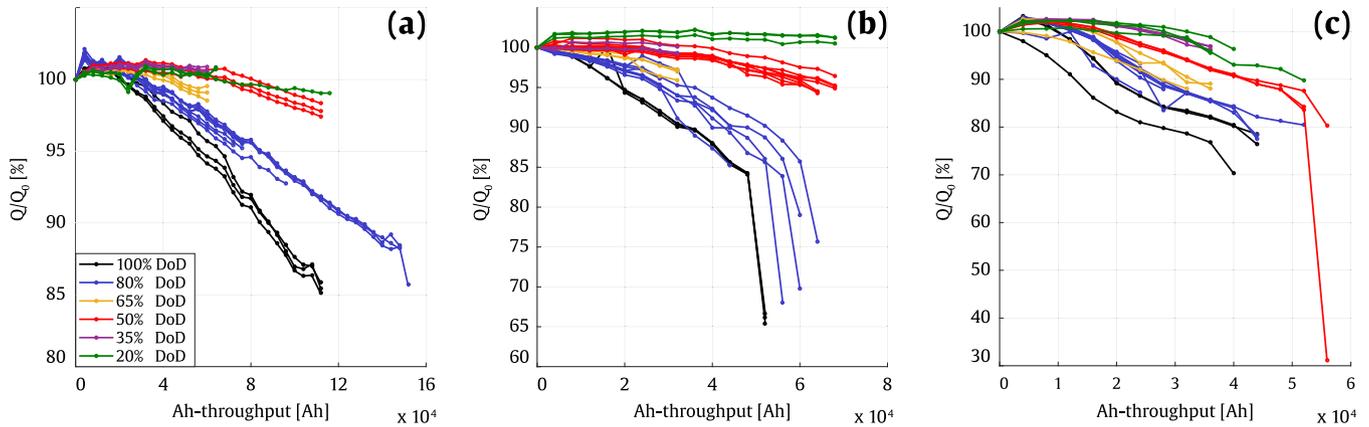

## Middle-SOC dependency

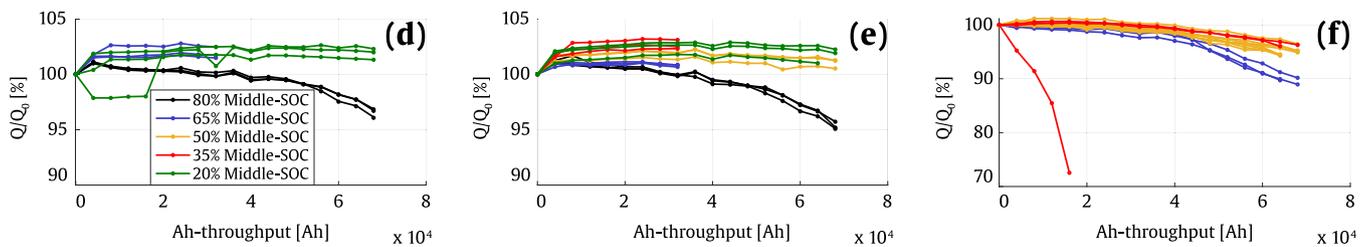

## Charging C-rate dependency

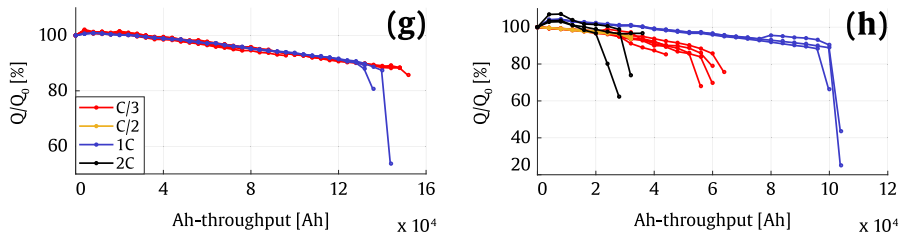

## Discharging C-rate dependency

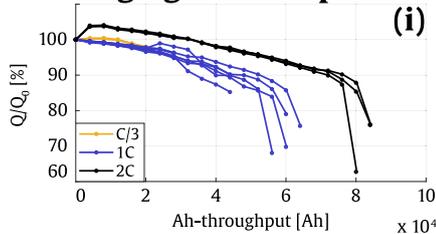

## Symmetric charging and discharging C-rate

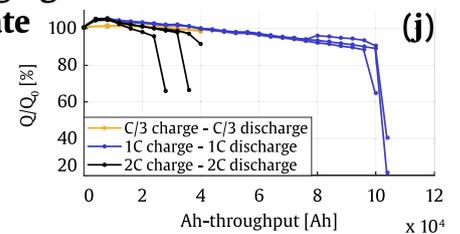

**Fig. B.1.** Normalised capacity (with initial value $Q_0$), obtained from the experimental static ageing tests at (a) 25°C, 50% middle-SOC, C/3 – 1C, and several DOD values, (b) 35°C, 50% middle-SOC, C/3 – 1C, and several DOD values, (c) 45°C, 50% middle-SOC, C/3 – 1C, and several DOD values, (d) 35°C, 10% DOD, C/3 – 1C, and several middle-SOC values, (e) 35°C, 20% DOD, C/3 – 1C, and several middle-SOC values, (f) 35°C, 50% DOD, C/3 – 1C, and several middle-SOC values, (g) 25°C, 80% DOD, 50% middle-SOC, 1C discharging rate, and several charging rate values, (h) 35°C, 80% DOD, 50% middle-SOC, 1C discharging rate, and several charging rate values, (i) 35°C, 80% DOD, 50% middle-SOC, C/3 charging rate, and several discharging rate values, and (j) 35°C, 80% DOD, 50% middle-SOC and several symmetric charging and discharging rate values.

In Section 7, the developed model was validated at dynamic operating conditions, and the ability of the model to learn directly from dynamic operating conditions was illustrated with 2 cells. In Fig. 8(a), the initial model overestimated the degradation trend, mainly due to a large cycling step at 100% DOD around 80000 Ah-throughput. In fact, the cells tested at 100% DOD in static cycling conditions observed increased capacity losses, compared to the cell tested at the same DOD but within a dynamic ageing test. A similar behaviour was already reported in [35], in which it was suggested that the dynamic character of the DOD stress-factor's profile may induce reduced ageing rates compared to static DOD profiles. This observation increases the interest of ageing models able to learn from the dynamic profiles observed after deployment, correcting this way the initial model trained with laboratory static ageing experiments. Anyway, this observation should be verified in further work, and the study should be extended involving more cells cycled at dynamic conditions.

Furthermore, the operating conditions collected from the battery-packs deployed in the real application are expressed in term of temperature, current and voltage time series. Therefore, algorithms should be developed to convert the collected time series data into stress-factors profiles (e.g. SOC estimation algorithms [36]), in order to input them to the models developed in this paper. Additionally, as the proposed models are trained under the supervised learning paradigm, the output of the model also needs to be estimated from real-world application, in order to obtain a complete training data involving both inputs and the corresponding output. The target training data could be obtained by i) periodical characterisation tests, if this is allowed by the application, or by ii) periodical SOH estimations, performed by dedicated algorithms, which is also another active research field for Li-ion batteries [37,38].

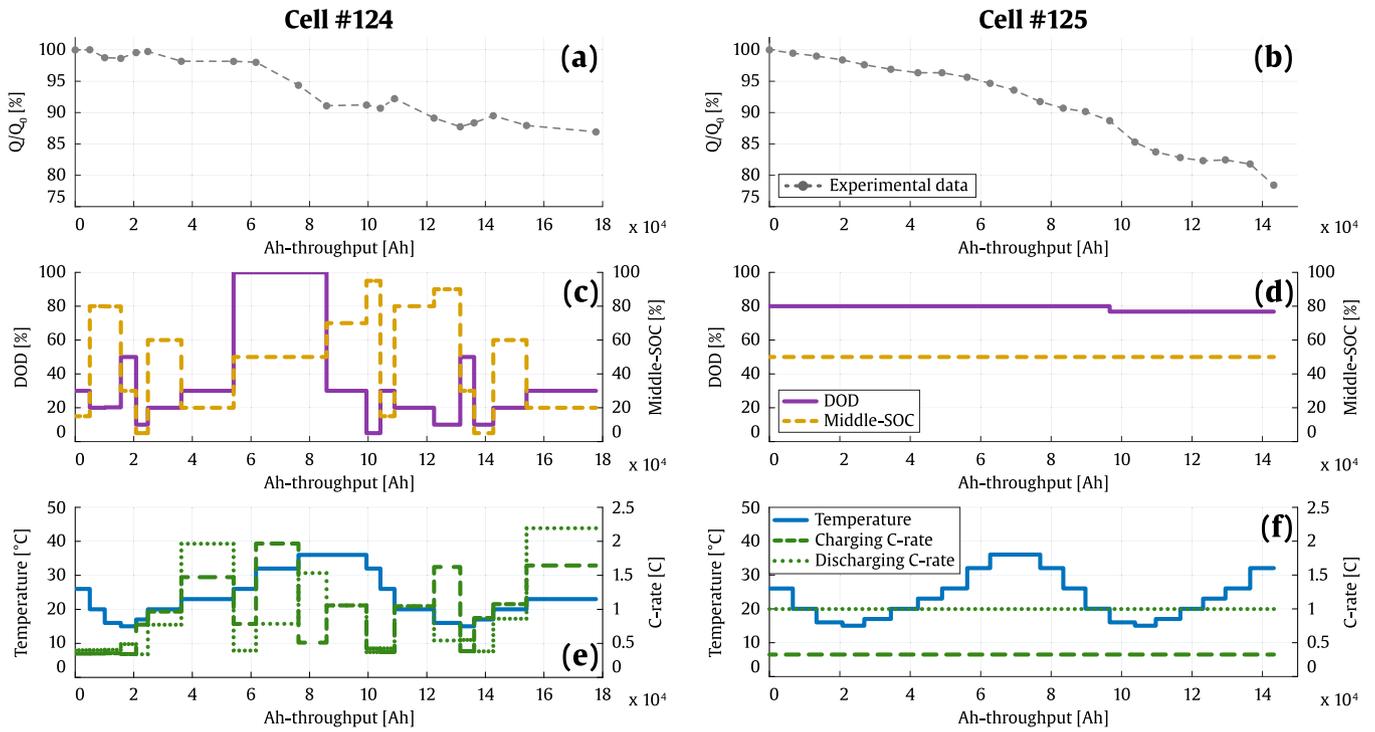

**Fig. B.2.** (a) Normalised capacity (with initial value $Q_0$), and the corresponding (c) DOD and middle-SOC, and (e) temperature and charging and discharging C-rate profiles, for the cell #124. (b) Normalised capacity (with initial value $Q_0$), and the corresponding (d) DOD and middle-SOC, and (f) temperature and charging and discharging C-rate profiles, for the cell #125.

**Table A1**
Cycle ageing tests matrix, for the tests at static ageing conditions.

| Temperature [°C] | | 25 | | 35 | | | | | | | 45 |
|---|---|---|---|---|---|---|---|---|---|---|---|
| C-Rate [C] (charge - discharge) | | C/3-1C | 1C-1C | C/3-C/3 | C/3-1C | C/3-2C | C/2-1C | 1C-1C | 2C-1C | 2C-2C | C/3-1C |
| DoD [%] | MidSOC [%] | | | | | | | | | | |
| 100 | 50 | 3 | | | 3 | | | | | | 3 |
| 80 | 50 | 8 | 3 | 3 | 8 | 3 | 3 | 3 | 3 | 3 | 8 |
| 65 | 50 | 3 | | | 3 | | | | | | 3 |
| 50 | 65 | | | | 3 | | | | | | |
|  | 50 | 3 | | | 8 | | | | | | 3 |
|  | 35 | | | | 3 | | | | | | |
| 35 | 50 | 3 | | | 3 | | | | | | 3 |
| 20 | 80 | | | | 3 | | | | | | |
|  | 65 | | | | 3 | | | | | | |
|  | 50 | 3 | | | 3 | | | | | | 3 |
|  | 35 | | | | 3 | | | | | | |
|  | 20 | | | | 3 | | | | | | |
| 10 | 80 | | | | 3 | | | | | | |
|  | 65 | | | | 3 | | | | | | |
|  | 20 | | | | 3 | | | | | | |

Finally, the deployment of the cycle ageing model developed in this study deserves a deeper discussion from the perspective of computational complexity. In fact, the logging of the current, voltage and temperature time series, as well as the extraction of the corresponding inputs variables requires memory and computation considerations. Furthermore, the inherent time and memory complexity of the GP is $O(n^3)$ and $O(n^2)$, and the required computations rapidly become prohibitive within the context of increasing training datasets. Within this context, two different approaches could be contemplated for the deployment of ageing models in real applications, considering the implementation of the models i) within the local hardware of each battery system, or ii) in an external data server (cloud server), connected to a fleet of battery systems (Fig. 10).

The first approach presents several issues related to the computational complexity of the ageing model. In fact, the above-mentioned $O(n^3)$ and $O(n^2)$ time and memory complexity of the GP questions its ability to be implemented within a device with limited computation resources, in the context of increasing training datasets. This implies that i) approximation methods of the GP algorithm [32] may be required once the training dataset becomes critically large, and ii) the model must be implemented in a hardware system presenting minimal requirements in terms of computation power. In case the model would be implemented in a BMS board, an oversizing of the system could be necessary, as well as an adequate coordination with the other tasks performed by the BMS (e.g. measurements, safety-related tasks, SOC estimations, communications, etc.). However, the first approach presents the advantage of avoiding the cost corresponding to the implementation of the data server and communication systems with the local devices. Furthermore, as most of the battery systems commercially available are oriented to a local operation and lack communication capabilities, it is likely that such an approach may be the more widely adopted in the short-term.

**Table B1**
Mean variance of the raw capacity curves, for all the cells tested at identical storage conditions (in [%²]).

| Temperature [°C] | | 25 | | | 35 | | | | | | | 45 |
|---|---|---|---|---|---|---|---|---|---|---|---|---|
| C-Rate [C] (charge - discharge) | | C/3-1C | 1C-1C | C/3-C/3 | C/3-1C | C/3-2C | C/2-1C | 1C-1C | 2C-1C | 2C-2C | | C/3-1C |
| DoD [%] | MidSOC [%] | | | | | | | | | | | |
| 100 | 50 | 0.26 | | | 0.41 | | | | | | | 12.82[6] |
| 80 | 50 | 0.11 | 0.03 | 0.01 | 0.27 | 9.29[2] | 0.13 | 7.94[3] | 69.89[4] | 40.88[5] | | 1.93 |
| 65 | 50 | 0.04 | | | 0.21 | | | | | | | 3.26 |
| 50 | 65 | | | | 0.41 | | | | | | | |
|  | 50 | 0.12 | | | 0.29 | | | | | | | 0.40 |
|  | 35 | | | | 74.06[1] | | | | | | | |
| 35 | 50 | 0.01 | | | 0.16 | | | | | | | 0.06 |
| 20 | 80 | | | | 0.05 | | | | | | | |
|  | 65 | | | | 0.01 | | | | | | | |
|  | 50 | 0.03 | | | 0.14 | | | | | | | 1.04 |
|  | 35 | | | | 0.15 | | | | | | | |
|  | 20 | | | | 0.37 | | | | | | | |
| 10 | 80 | | | | 0.03 | | | | | | | |
|  | 65 | | | | 0.23 | | | | | | | |
|  | 20 | | | | 1.18 | | | | | | | |

[1] This high variance value is induced by the cell #56, which depicts a clearly defective behaviour (isolated red curve in Fig. B.1 **(f)**, Appendix B).
[2] This high variance is induced by the sudden capacity drops observable in the black curves, Fig. B.1 **(i)**, Appendix B.
[3] These high variances are induced by the sudden capacity drops observable in the blue and black curves respectively, Fig. B.1 **(j)**, Appendix B.
[4] This high variance is induced by the sudden capacity drops observable in the black curves, Fig. B.1 **(h)**, Appendix B.
[5] These high variances are induced by the sudden capacity drops observable in the blue and black curves respectively, Fig. B.1 **(j)**, Appendix B.
[6] This high variance is induced by a shift between the different curves obtained at this testing conditions, Fig. B.1 **(c)**, Appendix B.

The second approach provides many advantages with respect to the first one. First, it allows a single ageing model to be fed in parallel with the data obtained from several battery systems, empowering it to learn simultaneously about the effect of a wide range of operating conditions. In this way, the learning process of the GP could be significantly accelerated, and the resulting ageing model could provide reliable predictions on a broader range of the operating conditions, compared to the different models evolving independently in the context of the first approach. Furthermore, the observation of the data collected from several battery systems allows to assess the quality and variability of the obtained data, from a statistical point of view. This also permits a correct modelling of the noise induced by the measurement sensors of the deployed system. From a long-term perspective, this second approach produces a database traducing the behaviour of the deployed systems in real applications. This information could be exploited for further objectives, e.g. to acquire a better knowledge about the deployed systems, carry out further modelling works, extract new insights for the development of new business models, etc. The main drawback of such an approach is the implementation cost of the data server and the communication systems. Moreover, the issue of the computational complexity of the GP model would be linked to the computation power of the server, which would determine the necessity to use approximation methods oriented to reduce the time and memory complexity of the GP models.

the Gaussian Process framework. The model presented 0.58% $MAE_{\Delta Q}$ and 1.04% $MAE_Q$ average prediction errors for 122 cells operating between 25°C-45°C, 20%-100% DOD, 20%-80% middle-SOC, C/3-2C charging C-rates and C/3-2C discharging C-rates, using only 26 cells tested at 9 cycling conditions for training.

The research works carried out in this paper with cycle ageing data corroborates the findings observed with calendar ageing data in the first paper of the series: i) due to their nonparametric character, GP-based models are capable to learn from progressively observed operating conditions; this makes the GP framework a suitable candidate to develop ageing models able to evolve and improve their performances even after deployment in real application, ii) isotropic kernel components are suitable to host the features corresponding to the different stress-factors, in so far as the battery operates within the limited range of the recommended operating conditions.

The sensitivity analysis shows that, for this dataset, the developed model tends to classify the influence of the stress-factors' variation on capacity loss in this order: 1/ temperature, 2/ DOD, 3/ charging C-rate, 4/ middle-SOC and 5/ discharging C-rate.

Furthermore, the discrepancies between the capacity loss induced by static DOD profiles and dynamic DOD profiles, observed in this dataset and reported in the literature, highlight the increased interest of ageing models capable to evolve after the deployment and learn from the dynamic profiles observed in real applications.

The first paper of the series provided a detailed description of a counterpart ageing model, focussed on the storage operation of the battery. With the cycle ageing model introduced in this second paper, an overall predictive tool is provided to predict the capacity loss of Li-ion batteries at both storage and cycling use-cases. Furthermore, the authors are currently working on the integration of both models for applications which are sequentially subject to calendar and cycle ageing. The results are planned to be described in a further publication, using experimental ageing data corresponding to EV driving load profiles, as well as power smoothing renewable energy integration profiles applied to second life batteries.

## 9. Conclusions

In this paper, a cycling capacity loss model was developed based on

## Acknowledgments

This investigation work was financially supported by ELKARTEK

**Table C1**
1. Results obtained with the models resulting from the different training cases, for the training, validation, targeted validation and all the cells, in terms of ΔQ and Q.

| | Capacity loss (ΔQ) | | | | | | | | | | | | Capacity (Q) | | | | | | | | | | | | |
|---|---|---|---|---|---|---|---|---|---|---|---|---|---|---|---|---|---|---|---|---|---|---|---|---|---|
| | Training | | | Validation | | | Targeted validation | | | All | | | Training | | | Validation | | | Targeted validation | | | All | | |
| | MAE | RMSE | CS | MAE | RMSE | CS | MAE | RMSE | CS | MAE | RMSE | CS | MAE | RMSE | CS | MAE | RMSE | CS | MAE | RMSE | CS | MAE | RMSE | CS |
| CASE 1 | 1.13 | 1.36 | 84.72 | 2.09 | 2.36 | 84.35 | 1.01 | 1.22 | 100.00 | 2.04 | 2.31 | 84.37 | 1.43 | 1.67 | 80.36 | 5.97 | 6.63 | 36.57 | 4.02 | 4.25 | 43.32 | 5.74 | 6.38 | 38.76 |
| CASE 2 | 0.94 | 1.16 | 87.60 | 1.72 | 1.97 | 70.26 | 1.72 | 1.98 | 71.54 | 1.66 | 1.91 | 71.56 | 1.56 | 1.74 | 80.71 | 4.91 | 5.44 | 26.89 | 4.89 | 5.40 | 23.10 | 4.66 | 5.16 | 30.93 |
| CASE 3 | 0.70 | 0.85 | 90.87 | 0.68 | 0.84 | 95.22 | 0.77 | 0.96 | 94.26 | 0.69 | 0.84 | 94.57 | 1.06 | 1.19 | 81.61 | 1.92 | 2.11 | 56.84 | 2.18 | 2.36 | 47.30 | 1.79 | 1.97 | 60.55 |
| CASE 4 | 0.68 | 0.84 | 90.65 | 0.55 | 0.70 | 94.02 | 0.61 | 0.77 | 93.12 | 0.58 | 0.73 | 93.26 | 1.12 | 1.23 | 77.75 | 1.02 | 1.16 | 82.01 | 1.06 | 1.19 | 80.81 | 1.04 | 1.17 | 81.05 |
| CASE 5 | 0.57 | 0.71 | 90.03 | 0.58 | 0.73 | 90.48 | 0.71 | 0.89 | 87.12 | 0.58 | 0.72 | 90.33 | 1.06 | 1.17 | 72.31 | 1.04 | 1.18 | 75.18 | 1.12 | 1.25 | 69.62 | 1.05 | 1.17 | 74.20 |
| CASE 6 | 0.66 | 0.82 | 88.14 | 0.53 | 0.66 | 93.56 | 0.66 | 0.79 | 100.00 | 0.60 | 0.75 | 90.62 | 1.09 | 1.32 | 68.96 | 1.09 | 1.26 | 78.77 | 0.74 | 1.00 | 99.31 | 1.16 | 1.29 | 73.45 |
| CASE 7 | 0.62 | 0.77 | 88.79 | 0.50 | 0.63 | 92.20 | 0.32 | 0.43 | 96.26 | 0.58 | 0.72 | 90.10 | 1.11 | 1.24 | 68.74 | 1.11 | 1.25 | 75.05 | 0.50 | 0.62 | 93.75 | 1.13 | 1.25 | 71.16 |
| CASE 8 | 0.61 | 0.75 | 88.84 | 0.50 | 0.62 | 91.93 | 0.28 | 0.34 | 96.96 | 0.57 | 0.71 | 89.87 | 1.14 | 1.22 | 70.09 | 1.14 | 1.26 | 70.76 | 0.37 | 0.44 | 92.58 | 1.12 | 1.23 | 70.32 |
| CASE 9 | 0.57 | 0.71 | 89.28 | 0.55 | 0.68 | 89.81 | 0.30 | 0.38 | 95.17 | 0.57 | 0.70 | 89.43 | 1.11 | 1.22 | 71.53 | 1.28 | 1.40 | 65.63 | 0.40 | 0.46 | 91.88 | 1.11 | 1.22 | 69.86 |
| CASE 10 | 0.56 | 0.69 | 90.37 | 0.64 | 0.77 | 89.39 | 0.24 | 0.29 | 100.00 | 0.57 | 0.71 | 90.16 | 1.04 | 1.14 | 71.51 | 1.60 | 1.76 | 58.47 | 0.40 | 0.54 | 93.24 | 1.16 | 1.27 | 68.80 |
| CASE 11 | 0.54 | 0.67 | 90.59 | 0.71 | 0.86 | 87.37 | 0.78 | 0.94 | 85.31 | 0.57 | 0.71 | 90.03 | 1.01 | 1.11 | 71.77 | 1.83 | 1.99 | 51.02 | 2.15 | 2.36 | 44.32 | 1.15 | 1.26 | 68.13 |
| CASE 12 | 0.56 | 0.69 | 90.11 | 0.65 | 0.77 | 91.00 | 0.67 | 0.79 | 90.63 | 0.57 | 0.70 | 90.24 | 1.03 | 1.13 | 71.12 | 2.03 | 2.25 | 44.80 | 2.74 | 3.07 | 25.58 | 1.18 | 1.30 | 67.18 |
| CASE 13 | 0.54 | 0.67 | 89.79 | 0.64 | 0.77 | 89.85 | 0.71 | 0.84 | 90.48 | 0.55 | 0.68 | 89.80 | 1.01 | 1.11 | 71.35 | 1.29 | 1.42 | 59.88 | 1.32 | 1.47 | 61.40 | 1.04 | 1.14 | 70.21 |
| CASE 14 | 0.54 | 0.67 | 89.94 | 0.61 | 0.75 | 89.65 | 0.61 | 0.75 | 89.65 | 0.54 | 0.67 | 89.92 | 1.00 | 1.10 | 71.97 | 1.29 | 1.40 | 59.37 | 1.29 | 1.40 | 59.37 | 1.02 | 1.12 | 71.02 |
| CASE 15 | 0.54 | 0.67 | 89.89 | 0.63 | 0.80 | 85.19 | 0.63 | 0.80 | 85.19 | 0.54 | 0.67 | 89.77 | 1.00 | 1.09 | 72.28 | 1.08 | 1.12 | 61.90 | 1.08 | 1.12 | 61.90 | 1.00 | 1.10 | 72.02 |
| CASE 16 | 0.54 | 0.67 | 89.74 | NaN | NaN | NaN | NaN | NaN | NaN | 0.54 | 0.67 | 89.74 | 1.00 | 1.09 | 71.97 | NaN | NaN | NaN | NaN | NaN | NaN | 1.00 | 1.09 | 71.97 |

(CICe2018 - Desarrollo de actividades de investigación fundamental estratégica en almacenamiento de energía electroquímica y térmica para sistemas de almacenamiento híbridos, KK-2018/00098) and EMAITEK Strategic Programs of the Basque Government. In addition, the research was undertaken as a part of ELEVATE project (EP/M009394/1) funded by the Engineering and Physical Sciences Research Council (EPSRC) and partnership with the WMG High Value Manufacturing (HVM) Catapult.

Authors would like to thank the FP7 European project Batteries 2020 consortium (grant agreement No. 608936) for the valuable battery ageing data provided during the project.

**Appendix A. Cycle ageing test matrix**

**Appendix B. Raw capacity data obtained from experimental cycle ageing tests, and variability of the resulting capacity curves**

**Appendix C. Results obtained with the models resulting from the different training cases**

**References**

[1] G. Zubi, R. Dufo-López, M. Carvalho, G. Pasaoglu, The lithium-ion battery: state of the art and future perspectives, Renew. Sustain. Energy Rev. 89 (2018) 292–308, https://doi.org/10.1016/j.rser.2018.03.002.
[2] S. Dhundhara, Y.P. Verma, A. Williams, Techno-economic analysis of the lithium-ion and lead-acid battery in microgrid systems, Energy Convers. Manag. 177 (2018) 122–142, https://doi.org/10.1016/j.enconman.2018.09.030.
[3] N. Nitta, F. Wu, J.T. Lee, G. Yushin, Li-ion battery materials: present and future, Mater. Today 18 (2015) 252–264, https://doi.org/10.1016/j.mattod.2014.10.040.
[4] M. Lucu, E. Martinez-Laserna, I. Gandiaga, K. Liu, H. Camblong, W.D. Widanage, J. Marco, Data-driven nonparametric Li-ion battery ageing model aiming at learning from real operation data – Part A: storage operation, J. Energy Storage 30 (2020), https://doi.org/10.1016/j.est.2020.101409.
[5] M. Lucu, E. Martinez-Laserna, I. Gandiaga, H. Camblong, A critical review on self-adaptive Li-ion battery ageing models, J. Power Sources 401 (2018) 85–101, https://doi.org/10.1016/j.jpowsour.2018.08.064.
[6] K.A. Severson, P.M. Attia, N. Jin, N. Perkins, B. Jiang, Z. Yang, M.H. Chen, M. Aykol, P.K. Herring, D. Fraggedakis, M.Z. Bazant, S.J. Harris, W.C. Chueh, R.D. Braatz, Data-driven prediction of battery cycle life before capacity degradation, Nat. Energy. 4 (2019) 383–391, https://doi.org/10.1038/s41560-019-0356-8.
[7] R.R. Richardson, M.A. Osborne, D.A. Howey, Gaussian process regression for forecasting battery state of health, J. Power Sources 357 (2017) 209–219, https://doi.org/10.1016/j.jpowsour.2017.05.004.
[8] R.R. Richardson, M.A. Osborne, D.A. Howey, Battery health prediction under generalized conditions using a Gaussian process transition model, J. Energy Storage 23 (2019) 320–328, https://doi.org/10.1016/j.est.2019.03.022.
[9] D. Liu, J. Pang, J. Zhou, Y. Peng, M. Pecht, Prognostics for state of health estimation of lithium-ion batteries based on combination Gaussian process functional regression, Microelectron. Reliab. 53 (2013) 832–839, https://doi.org/10.1016/j.microrel.2013.03.010.
[10] Y. He, J.-N. Shen, J.-F. Shen, Z.-F. Ma, State of health estimation of lithium-ion batteries: a multiscale Gaussian process regression modeling approach, AIChE J. 61 (2015) 1589–1600, https://doi.org/10.1002/aic.14760.
[11] L. Li, P. Wang, K.-H. Chao, Y. Zhou, Y. Xie, Remaining useful life prediction for lithium-ion batteries based on gaussian processes mixture, PLoS One 11 (2016) e0163004, , https://doi.org/10.1371/journal.pone.0163004.
[12] D. Yang, X. Zhang, R. Pan, Y. Wang, Z. Chen, A novel Gaussian process regression model for state-of-health estimation of lithium-ion battery using charging curve, J. Power Sources 384 (2018) 387–395, https://doi.org/10.1016/j.jpowsour.2018.03.015.
[13] J. Yu, State of health prediction of lithium-ion batteries: multiscale logic regression and Gaussian process regression ensemble, Reliab. Eng. Syst. Saf. 174 (2018) 82–95, https://doi.org/10.1016/j.ress.2018.02.022.
[14] Y. Peng, Y. Hou, Y. Song, J. Pang, D. Liu, Lithium-ion battery prognostics with hybrid gaussian process function regression, Energies 11 (2018) 1420, https://doi.org/10.3390/en11061420.
[15] K. Liu, Y. Li, X. Hu, M. Lucu, D. Widanalage, Gaussian process regression with automatic relevance determination kernel for calendar aging prediction of lithium-

ion batteries, IEEE Trans. Ind. Inf. (2019) 1–1, doi:10.1109/TII.2019.2941747.
[16] E. Redondo-iglesias, Calendar and cycling ageing combination of batteries in electric vehicles, Microelectron. Reliab. 88–90 (2018) 1212–1215, https://doi.org/10.1016/j.microrel.2018.06.113.
[17] E. Redondo-iglesias, Étude du vieillissement des batteries lithium-ion dans les applications ″ véhicule électrique ″ : combinaison des effets de vieillissement calendaire et de cyclage, Université de Lyon (2017), https://tel.archives-ouvertes.fr/tel-01668529/document.
[18] E. Sarasketa-Zabala, E. Martinez-Laserna, M. Berecibar, I. Gandiaga, L.M. Rodriguez-Martinez, I. Villarreal, Realistic lifetime prediction approach for Li-ion batteries, Appl. Energy 162 (2016) 839–852, https://doi.org/10.1016/j.apenergy.2015.10.115.
[19] L. Zhou, S. Pan, J. Wang, A.V. Vasilakos, Machine learning on big data: Opportunities and challenges, Neurocomputing 237 (2017) 350–361, https://doi.org/10.1016/j.neucom.2017.01.026.
[20] A.J. Warnecke, Degradation Mechanisms in NMC-Based Lithium-Ion Batteries (2017).
[21] X. Yang, C. Wang, Understanding the trilemma of fast charging, energy density and cycle life of lithium-ion batteries, J. Power Sources 402 (2018) 489–498, https://doi.org/10.1016/j.jpowsour.2018.09.069.
[22] D. Li, H. Li, D. Danilov, L. Gao, J. Zhou, R.A. Eichel, Y. Yang, P.H.L. Notten, Temperature-dependent cycling performance and ageing mechanisms of C6/LiNi1/3Mn1/3Co1/3O2 batteries, J. Power Sources 396 (2018) 444–452, https://doi.org/10.1016/j.jpowsour.2018.06.035.
[23] R.D. Deshpande, D.M. Bernardi, Modeling solid-electrolyte interphase (SEI) fracture: coupled mechanical/chemical degradation of the lithium ion battery, J. Electrochem. Soc. 164 (2017) A461–A474, https://doi.org/10.1149/2.0841702jes.
[24] A. Marongiu, M. Roscher, D.U. Sauer, Influence of the vehicle-to-grid strategy on the aging behavior of lithium battery electric vehicles, Appl. Energy 137 (2015) 899–912, https://doi.org/10.1016/j.apenergy.2014.06.063.
[25] I. Laresgoiti, S. Käbitz, M. Ecker, D.U. Sauer, Modeling mechanical degradation in lithium ion batteries during cycling: solid electrolyte interphase fracture, J. Power Sources 300 (2015) 112–122, https://doi.org/10.1016/j.jpowsour.2015.09.033.
[26] M. Ecker, N. Nieto, S. Käbitz, J. Schmalstieg, H. Blanke, A. Warnecke, D.U. Sauer, Calendar and cycle life study of Li(NiMnCo)O2-based 18650 lithium-ion batteries, J. Power Sources 248 (2014) 839–851, https://doi.org/10.1016/j.jpowsour.2013.09.143.
[27] M. Dubarry, N. Qin, P. Brooker, Calendar aging of commercial Li-ion cells of different chemistries – A review, Curr. Opin. Electrochem. 9 (2018) 106–113, https://doi.org/10.1016/j.coelec.2018.05.023.
[28] A. Mukhopadhyay, B.W. Sheldon, Deformation and stress in electrode materials for Li-ion batteries, Prog. Mater. Sci. 63 (2014) 58–116, https://doi.org/10.1016/j.pmatsci.2014.02.001.
[29] R. Deshpande, M. Verbrugge, Y.-T. Cheng, J. Wang, P. Liu, Battery cycle life prediction with coupled chemical degradation and fatigue mechanics, J. Electrochem. Soc. 159 (2012) A1730–A1738, https://doi.org/10.1149/2.049210jes.
[30] T. Waldmann, B.-I. Hogg, M. Wohlfahrt-Mehrens, Li plating as unwanted side reaction in commercial Li-ion cells – A review, J. Power Sources 384 (2018) 107–124, https://doi.org/10.1016/j.jpowsour.2018.02.063.
[31] S. Ahmed, I. Bloom, A.N. Jansen, T. Tanim, E.J. Dufek, A. Pesaran, A. Burnham, R.B. Carlson, F. Dias, K. Hardy, M. Keyser, C. Kreuzer, A. Markel, A. Meintz, C. Michelbacher, M. Mohanpurkar, P.A. Nelson, D.C. Robertson, D. Scoffield, M. Shirk, T. Stephens, R. Vijayagopal, J. Zhang, Enabling fast charging – A battery technology gap assessment, J. Power Sources 367 (2017) 250–262, https://doi.org/10.1016/j.jpowsour.2017.06.055.
[32] C.E. Rasmussen, C.K.I. Williams, Gaussian Processes for Machine Learning, MIT Press, 2006, http://www.gaussianprocess.org/gpml/chapters/RW.pdf.
[33] D. Duvenaud, J.R. Lloyd, R. Grosse, J.B. Tenenbaum, Z. Ghahramani, Structure discovery in nonparametric regression through compositional kernel search, Proc. 30th Int. Conf. Mach. Learn. PMLR 28 (3) (2013) 1166–1174 http://arxiv.org/abs/1302.4922.
[34] J.Q. Candela, A. Girard, J. Larsen, C.E. Rasmussen, Propagation of uncertainty in Bayesian kernel models - application to multiple-step ahead forecasting, 2003 IEEE Int. Conf. Acoust. Speech, Signal Process. 2003. Proceedings. (ICASSP '03), IEEE, 2003II-701–4, doi:10.1109/ICASSP.2003.1202463.
[35] E. Sarasketa-Zabala, I. Gandiaga, E. Martinez-Laserna, L.M. Rodriguez-Martinez, I. Villarreal, Cycle ageing analysis of a LiFePO4/graphite cell with dynamic model validations: Towards realistic lifetime predictions, J. Power Sources 275 (2015) 573–587, https://doi.org/10.1016/j.jpowsour.2014.10.153.
[36] Z. Li, J. Huang, B.Y. Liaw, J. Zhang, On state-of-charge determination for lithium-ion batteries, J. Power Sources 348 (2017) 281–301, https://doi.org/10.1016/j.jpowsour.2017.03.001.
[37] M. Berecibar, I. Gandiaga, I. Villarreal, N. Omar, J. Van Mierlo, P. Van Den Bossche, Critical review of state of health estimation methods of Li-ion batteries for real applications, Renew. Sustain. Energy Rev. 56 (2016) 572–587, https://doi.org/10.1016/j.rser.2015.11.042.
[38] S. Khaleghi, Y. Firouz, M. Berecibar, J. Van Mierlo, P. Van den Bossche, Ensemble gradient boosted tree for soh estimation based on diagnostic features, Energies 13 (2020) 1262, https://doi.org/10.3390/en13051262.